\documentstyle[aps,eqsecnum,preprint]{revtex}

\begin{document}

\tightenlines

\preprint{hep-ph/0005224}

\title{Nonequilibrium Quantum Dynamics of Second Order Phase Transitions}

\author{Sang Pyo Kim\footnote{E-mail:
sangkim@ks.kunsan.ac.kr; spkim@phys.ualberta.ca}\footnote{On leave
to Theoretical Physics Institute, Department of Physics,
University of Alberta, Edmonton, Alberta, Canada T6G 2J1}}
\address{Department of Physics\\ Kunsan National University\\Kunsan 573-701,
Korea}
\author{Chul H. Lee\footnote{E-mail: chlee@hepth.hanyang.ac.kr}}
\address{Department of Physics\\
Hanyang University\\ Seoul 133-791, Korea}

\date{\today}

\maketitle
\begin{abstract}
We use the so-called Liouville-von Neumann (LvN) approach to study
the nonequilibrium quantum dynamics of time-dependent second order
phase transitions. The LvN approach is a canonical method that
unifies the functional Schr\"{o}dinger equation for the quantum
evolution of pure states and the LvN equation for the quantum
description of mixed states of either equilibrium or
nonequilibrium. As nonequilibrium quantum mechanical systems we
study a time-dependent harmonic and an anharmonic oscillator and
find the exact Fock space and density operator for the harmonic
oscillator and the nonperturbative Gaussian Fock space and density
operator for the anharmonic oscillator. The density matrix and the
coherent, thermal and coherent-thermal states are found in terms
of their classical solutions, for which the effective Hamiltonians
and equations of motion are derived. The LvN approach is further
extended to quantum fields undergoing time-dependent second order
phase transitions. We study an exactly solvable model with a
finite smooth quench and find the two-point correlation functions.
Due to the spinodal instability of long wavelength modes the
two-point correlation functions lead to the $t^{1/4}$-scaling
relation during the quench and the Cahn-Allen scaling relation
$t^{1/2}$ after the completion of quench. Further, after the
finite quench the domain formation shows a time-lag behavior at
the cubic power of quench period. Finally we study the
time-dependent phase transition of a self-interacting scalar
field.
\end{abstract}


\section{Introduction}

A system can interact directly with an environment to make its
coupling parameters depend explicitly on time. Even the effective
coupling parameters of a subsystem of a closed system, though
conserved as a whole, may depend implicitly on time through an
interaction with the rest of the system. The characteristic
feature of these open systems is that their effective coupling
parameters depend explicitly or implicitly on time. Therefore the
genuine understanding of these systems requires the real-time
processes from their initial conditions. Of a particular interest
are the systems undergoing phase transitions. When a system cools
down through an interaction with an environment, it may undergo a
phase transition and its coupling parameters depend explicitly on
time. Similarly, matter fields in the expanding early Universe
undergo phase transitions through the interaction with gravity.

Phase transitions are one of the most physically important
phenomena in nature and have wide applications from condensed
matter physics, particle physics and even to cosmology. The Kibble
mechanism explains formation of topological defects in symmetry
breaking phase transitions \cite{kibble}. The kinetic process by
Zurek has revealed a new feature of symmetry breaking phase
transitions \cite{zurek}. The dynamics of phase transitions and
formation of topological defects since then have become an
important tool in variety of phenomena in condensed matter physics
\cite{bray}. It is also widely accepted that symmetry breaking
phase transitions in the early Universe are inevitable for the
structure formation of the present Universe \cite{vilenkin}. There
has been an attempt through laboratory experiments to investigate
the process of structure formation in the early stage of Universe
\cite{zurek2}. In QCD the quark-antiquark condensate breaks chiral
symmetry when the temperature of quark-gluon plasma is lowered
\cite{wilczek,bellac}.

However, the most difficult and subtle facet of phase transitions
is to understand its dynamics and the formation process of
topological defects. As emphasized above, systems do become out of
equilibrium in general during phase transitions because their
coupling parameters depend explicitly on time through the
interaction with an environment (heat bath). Hence the
nonequilibrium dynamics of phase transitions should differ
significantly from the equilibrium dynamics. In finite temperature
field theory one obtains the effective action for the system in a
thermal equilibrium or quasi-equilibrium by calculating its
quantum fluctuations about a vacuum \cite{dolan}. But as the phase
transition proceeds, the fluctuations grow and the stability is
lost. The effective action, when extrapolated to the phase
transition regime in a literal sense, has a complex value, the
imaginary part of which is related with the decay rate of the
false vacuum \cite{weinberg}. In this sense finite temperature
field theory can not be directly applied to study symmetry
breaking phase transitions.

Schwinger and Keldysh introduced the closed time-path integral to
treat properly the quantum evolution of quantum fields out of
equilibrium from their initial thermal equilibrium
\cite{schwinger}. Since then the closed time-path integral method
has been developed and applied to many related problems
\cite{landsman}. Recently the closed time-path integral method has
been employed to study the nonequilibrium dynamics of second order
phase transitions
\cite{boyanovsky,bowick,calzetta,boyanovsky2,rivers}. Another
method is the functional Schr\"{o}dinger-picture approach, in
which the evolution of quantum states is found for explicitly
time-dependent Hamiltonians \cite{freese}. The large $N$-expansion
method \cite{cooper,destri} and the mean-field or Hartree-Fock
method \cite{ringwald,cormier1,cormier2} are used in conjunction
with either the Schwinger-Keldish or functional Schr\"{o}dinger
method. Still another methods are the time-dependent variational
principle \cite{balian}, the generating function for correlation
functions \cite{wetterich}, and thermal field theory
\cite{umezawa}.

The purpose of this paper is two-fold. In the first part of this
paper we elaborate and establish the recently introduced
Liouville-von Neumann (LvN) approach so that it can readily be
applied to nonequilibrium dynamics. In the second part we apply
the approach to the systems undergoing time-dependent second order
phase transitions. The LvN approach is a canonical method that
unifies the functional Schr\"{o}dinger equation for the quantum
evolution of pure states and the LvN equation for the quantum
description of mixed states of either equilibrium or
nonequilibrium. One of the advantages is that one can make use of
the well-known techniques of quantum mechanics and quantum
many-particle systems. It is based on the observation by Lewis and
Riesenfeld \cite{lewis} that the quantum LvN equation, which is
originally used to define the density operator for a mixed state,
can also be used to find the exact pure states of a time-dependent
harmonic oscillator. This observation makes it possible to find
not only the mixed state but also the pure state of a
time-dependent system. This LvN approach has been developed and
applied to quantum fields in an expanding
Friedmann-Robertson-Walker universe \cite{kim,kim2,kim3} and to
open boson and fermion systems \cite{kim4}.

In this paper we particularly focus on the model systems whose
coupling parameters change signs during the evolution and
emphasize the role these systems playing in the second order phase
transition. In the case of a time-dependent harmonic oscillator or
an ensemble of such oscillators with a positive time-dependent
frequency squared, the Fock space consists of the number states
which are the exact quantum states of the Schr\"{o}dinger equation
\cite{lewis,kim5}. The density operator is constructed in terms of
the annihilation and creation operators that satisfy the LvN
equation \cite{kim2}. Hence the coherent, thermal and
coherent-thermal states are found rather straightforwardly
according to the standard technique of quantum mechanics. We
further show that the same construction of the Fock space and
density operator still holds for the time-dependent harmonic
oscillator with a sign changing frequency squared. By studying
some analytically solvable models we investigate how the
instability grows. Another technical strong point of the LvN
approach is that it can also be used for a time-dependent
anharmonic oscillator. At the leading order the LvN approach is
equipped with the time-dependent annihilation and creation
operators, the vacuum state of which is already the time-dependent
Gaussian state that minimizes the Dirac action \cite{balian,kim6}.
The LvN approach thus provides one with a nonperturbative quantum
description for the time-dependent anharmonic oscillator, too. We
find the coherent, thermal and coherent-thermal states for the
anharmonic oscillator with a polynomial potential and study the
dynamics from their effective actions.

As field models for the second order phase transition, we consider
first a free massive and then a self-interacting scalar field, the
mass of which changes the sign in a finite time period through an
external interaction. By studying analytically the exactly
solvable model of the free scalar field, we show how the
instability of long wavelength modes grows in time. The two-point
thermal correlation function is expressed in terms of the
classical solution for each mode that is already found in terms of
a well-known function. The domain sizes are evaluated analytically
by the steepest descent method and are shown to grow as $t^{1/4}$
during the quench and as the Cahn-Allen scaling relation $t^{1/2}$
after the completion of quench. Remarkably, the Cahn-Allen scaling
relation shows a time-lag given by the cubic power of the quench
period, which is absent in the instantaneous quench model
\cite{boyanovsky,bowick}. However, both the instantaneous and the
finite quench models have the same scaling relation given by the
classical Cahn-Allen equation \cite{bray}, confirming the result
of the instantaneous quench model \cite{boyanovsky,bowick}. The
free scalar field model describes only the stage of spinodal
instability from the unstable local maximum to the spinodal line.
In the self-interacting field model, the back-reaction of an
interacting term contributes positively to the frequencies of long
wavelength modes and shuts off the exponential growth of
instability after crossing the spinodal line at sufficiently later
times.

The organization of this paper is as follows. It mainly consists
of two parts: in the first part from Sec. II to Sec. V the LvN
approach is elaborated to be applicable to phase transitions, and
in the second part of Secs. VI and VII it is then applied to the
second order phase transitions. In Sec. II we review the LvN
approach to time-dependent quantum systems. In Sec. III we study
the time-dependent harmonic oscillator and find the exact Fock
space and the density operator. The density matrix is found and
the nonequilibrium quantum dynamics is studied for the coherent,
thermal and coherent-thermal states. In Sec. IV the LvN approach
is applied to time-dependent inverted oscillators as a toy model
for the second order phase transition. In Sec. V we extend the LvN
approach to time-dependent anharmonic oscillators with polynomial
potentials. The LvN approach leads to the nonperturbative Gaussian
state at the leading order. We also show that the coherent state
of the LvN approach recovers the result from the time-dependent
mean-field or Hartree-Fock method. In Sec. VI we study an exactly
solvable model of a free scalar field, which has a time-dependent
mass coupling parameter and undergoes smoothly the second order
phase transition for a finite quench period. The two-point thermal
correlation function is analytically evaluated during the quench
and after the completion of quench, and the scaling relations for
the domain size are found. In Sec. VII we study a self-interacting
scalar field that undergoes the time-dependent second order phase
transition.

\section{Liouville-von Neumann (LvN) Approach}

In this section we briefly review but emphasize the underlying
assumptions of the LvN approach to time-dependent quantum systems
introduced in Ref. \cite{kim}. A time-dependent system can not
remain in the initial equilibrium or the instantaneous
quasi-equilibrium because the density operator
\begin{equation}
\rho_{\rm H} = \frac{1}{Z_{\rm H} (t)} e^{- \beta \hat{H} (t)},
\label{ham den}
\end{equation}
where $\hat{H} (t)$ is a time-dependent Hamiltonian operator and
$Z_{\rm H}$ is the partition function, does not satisfy the
quantum LvN equation. This means that though the system starts in
the initial thermal equilibrium, its final state can be far away
from the initial one. Even an initial pure state evolves toward a
final one which differs drastically from the initial one, and
leads, for instance, to particle production
\cite{parker,dewitt,birrel,kim7}. The Fock or Hilbert space of
such time-dependent system of a quantum field transforms unitarily
inequivalently so that even the initial vacuum state evolves to a
superposition of particle states at final times. Thus the
nonequilibrium system described by time-dependent quantum
Hamiltonian follows the evolution of a mixed state that is out of
equilibrium and is characterized by time-dependent processes. To
properly describe the nonequilibrium evolution we shall adopt two
assumptions.

First, from a microscopic point of view we assume that the
nonequilibrium system obeys the quantum law, {\it i.e.} the
time-dependent Schr\"{o}dinger or Tomonaga-Schwinger equation
\begin{equation}
i \hbar \frac{\partial}{\partial t} \vert \Psi (t) \rangle =
\hat{H} (t) \vert \Psi (t) \rangle. \label{ts eq}
\end{equation}
Here $\hat{H} (t)$ is the time-dependent Hamiltonian of the
system. This assumption is physically well-grounded since all the
individual constituents of the system should obey the quantum law
and the system as a whole should still obey the quantum law
provided that all the interactions among the individuals are
properly taken into account.

Second, from a statistical point of view it is assumed that even
the nonequilibrium system obeys the time-dependent quantum
Liouville-von Neumann (LvN) equation
\begin{equation}
i \hbar \frac{\partial}{\partial t} \hat{\rho} (t) + [ \hat{\rho}
(t), \hat{H} (t) ] = 0. \label{den eq}
\end{equation}
The LvN equation has been used to find the density operator for
equilibrium systems that are stationary. Now the nonequilibrium
system with their time-dependent coupling parameters follows the
same equation, so the density operator (\ref{ham den}) that is
directly defined in terms of the Hamiltonian itself does not
necessarily satisfy the equation. In Sec. III we shall see how
much the density operator satisfying Eq. (\ref{den eq}) deviates
from the instantaneous density operator (\ref{ham den}) for
time-dependent harmonic oscillators.

In the context of quantum mechanics a powerful canonical method
was developed by Lewis and Riesenfeld \cite{lewis}. They observed
that any operator $\hat{\cal O} (t)$ satisfying the quantum LvN
equation
\begin{equation}
i \hbar \frac{\partial}{\partial t} \hat{{\cal O}} (t) + [
\hat{{\cal O}} (t), \hat{H} (t)] = 0, \label{ln eq}
\end{equation}
can also be used to find the exact quantum states of Eq. (\ref{ts
eq}). In fact, any exact quantum state is given by
\begin{equation}
\vert \Psi (t) \rangle = \sum_{n} c_n e^{\frac{i}{\hbar} \gamma_n
(t)} \vert \lambda_n, t \rangle. \label{lin sup}
\end{equation}
where
\begin{eqnarray}
\hat{\cal O} (t) \vert \lambda_n, t \rangle = \lambda_{n} \vert
\lambda_n, t \rangle, \nonumber\\ \gamma_n (t) = \int dt \langle
\lambda_n, t \vert \Bigl( i \hbar \frac{\partial}{\partial t} -
\hat{H} (t) \Bigr) \vert \lambda_n, t \rangle. \label{phase}
\end{eqnarray}

Another useful property of LvN approach is the linearity of the
LvN equation. In fact, a product $\hat{\cal O}_1 (t) \hat{\cal
O}_2 (t)$ satisfies Eq. (\ref{ln eq}) whenever $\hat{\cal O}_1
(t)$ and $\hat{\cal O}_2 (t)$ satisfy Eq. (\ref{ln eq}). Therefore
it holds that any analytic functional $F[{\cal O} (t)]$ satisfies
Eq. (\ref{ln eq}) provided that ${\cal O} (t)$ satisfies the same
equation. In particular, we can still use $\hat{\cal O} (t)$ to
define the density operator for the time-dependent system
\begin{equation}
\hat{\rho}_{\cal O} (t) = \frac{e^{- \beta \hat{\cal O}
(t)}}{Z_{\cal O} (t)}, \quad Z_{\cal O} (t) = {\rm Tr} [e^{- \beta
\hat{\cal O} (t)}].
\end{equation}
Here $\beta$ is a free parameter and will be identified with the
inverse temperature only for the equilibrium system, in which the
Hamiltonian itself satisfies Eq. ({\ref{ln eq}) and is used to
define the standard density operator (\ref{ham den}).

\section{Time-Dependent Oscillator}

As a simple nonequilibrium system we begin with an ensemble of
time-dependent harmonic oscillators. This system carries more
meaning than being merely a toy model because most of systems with
some exception such as the massless $\Phi^4$-theory, either in
equilibrium or in nonequilibrium, can be approximated by a
quadratic Hamiltonian around zero-force points, stable or
unstable. In the technical aspect the time-dependent oscillator
can be exactly solved in terms of its classical solution. We now
focus on a general oscillator with a time-dependent mass and
frequency squared
\begin{equation}
\hat{H} (t) = \frac{1}{2m(t)} \hat{p}^2 + \frac{m(t)}{2}
\omega^2(t) \hat{q}^2, \label{osc}
\end{equation}
where $\omega^2(t)$ is allowed to change the sign during a phase
transition. The LvN approach will be employed below to find the
exact Fock space and to construct the coherent, thermal and
coherent-thermal states.

\subsection{Fock Space}

The key idea of the LvN approach to the Hamiltonian (\ref{osc}) is
to require a pair of operators \cite{kim,kim5}
\begin{eqnarray}
\hat{a} (t) &=&  i \Bigl(u^* (t) \hat{p} - m(t) \dot{u}^*
(t)\hat{q} \Bigr), \nonumber\\ \hat{a}^{\dagger} (t) &=&  - i
\Bigl(u (t) \hat{p} - m(t) \dot{u} (t) \hat{q} \Bigr),
\label{an-cr}
\end{eqnarray}
to satisfy the LvN equation (\ref{ln eq}). This results in the
classical equation of motion for the complex auxiliary variable
$u$
\begin{equation}
\ddot{u} (t) + \frac{\dot{m}(t)}{m(t)} \dot{u} (t) + \omega^2 (t)
u(t) = 0. \label{cl sol}
\end{equation}
Note that these operators depend explicitly on time through $u(t)$
and are hermitian conjugate to each other. Further, these
operators can be made the annihilation and creation operators with
the standard commutation relation for all the times
\begin{equation}
[\hat{a} (t), \hat{a}^{\dagger} (t)] = 1. \label{com rel}
\end{equation}
The above commutation relation is guaranteed by the wronskian
condition
\begin{equation}
\hbar m(t) \Bigl(\dot{u}^* (t) u(t) - \dot{u} (t) u^* (t) \Bigr) =
i. \label{boun con}
\end{equation}
A comment is in order. Nothing prevents one from using these
operators for an inverted time-dependent oscillator as far as Eq.
(\ref{boun con}) is satisfied. The inverted oscillator will be
treated in detail in Sec. IV in the context of second order phase
transitions.

>From the argument in Sec. II, using $\hat{a}(t)$ and
$\hat{a}^{\dagger}(t)$ one may construct two particular operators
that also satisfy Eq. (\ref{ln eq}): the number and the density
operator. By defining the number operator in the usual way
\begin{equation}
\hat{N} (t) = \hat{a}^{\dagger} (t) \hat{a} (t),
\end{equation}
one finds the Fock space consisting of the time-dependent number
states
\begin{equation}
\hat{N} (t) \vert n, t \rangle = n \vert n, t \rangle.\label{num
st}
\end{equation}
The vacuum state is the one that is annihilated by $\hat{a} (t)$
and the $n$th number state is obtained by applying
$\hat{a}^{\dagger}(t)$ $n$-times to the vacuum state:
\begin{eqnarray}
\hat{a} (t) \vert 0, t \rangle = 0, \nonumber\\ \vert n, t \rangle
= \frac{(\hat{a}^{\dagger} (t))^n}{\sqrt{n!}} \vert 0, t \rangle.
\end{eqnarray}
In the coordinate representation the number state is given by (see
Appendix A)
\begin{equation}
\Psi_n (q, t) = \Biggl(\frac{1}{2 \pi \hbar^2 u^* (t) u (t)}
\Biggr)^{1/4} \frac{1}{\sqrt{2^n n!}} \Biggl(\frac{u(t)}{u^* (t)}
\Biggr)^n H_n \Biggl(\frac{q}{\sqrt{2 \hbar^2 u^*(t) u (t)}}
\Biggr) \exp \Biggl[\frac{i}{2}\frac{m}{\hbar}
\frac{\dot{u}^*(t)}{u^*(t)} q^2 \Biggr], \label{har wav}
\end{equation}
where $H_n$ is the Hermite polynomial.

Equation (\ref{an-cr}) can be inverted to yield the position and
momentum operators
\begin{eqnarray}
\hat{q} = \hbar \Bigl( u (t) \hat{a}(t) + u^* (t)
\hat{a}^{\dagger}\Bigr), \nonumber\\ \hat{p} = \hbar m(t)
\Bigl(\dot{u} (t) \hat{a}(t) + \dot{u}^* (t)
\hat{a}^{\dagger}\Bigr).
\end{eqnarray}
Hence the expectation value of the position and momentum with
respect to each number state vanishes
\begin{equation}
\langle n, t \vert \hat{q} \vert n, t \rangle = \langle n, t \vert
\hat{p} \vert n, t \rangle = 0.
\end{equation}
The only nonvanishing expectation values come from even powers of
the position or momentum. The quadratic power of the position and
momentum has the expectation values
\begin{eqnarray}
\langle n, t \vert \hat{q}^2 \vert n, t \rangle &=&  \hbar^2 u^*
(t) u(t) ( 2n + 1), \nonumber\\ \langle n, t \vert \hat{p}^2 \vert
n, t \rangle &=&  \hbar^2 m^2 (t) \dot{u}^* (t) \dot{u}(t) ( 2n +
1), \nonumber\\
 \langle n, t \vert (\hat{q} \hat{p} + \hat{p} \hat{q}) \vert n,
t \rangle  &=&  \hbar^2 m (t) \Bigl(\dot{u}^* (t) u (t) + u^* (t)
\dot{u} (t) \Bigr) (2 n + 1).
\end{eqnarray}
The Hamiltonian thus has the expectation value
\begin{equation}
H_{\rm n} (t) = \langle n, t \vert \hat{H} (t) \vert n, t \rangle
= \frac{\hbar^2}{2} m(t) \Bigl[ \dot{u}^* (t) \dot{u} (t) +
\omega^2 (t) u^* (t) u(t) \Bigr] (2 n + 1). \label{num ex}
\end{equation}
The prominent advantages of using $\hat{a}^{\dagger} (t)$ and
$\hat{a}(t)$ appear when one tries, by using the standard
technique of quantum mechanics, to construct other quantum states
such as the coherent, thermal and coherent-thermal states.

\subsection{Coherent State}

The coherent state is a particularly useful quantum state in
quantum field theory for phase transitions or nonequilibrium
dynamics. It may be used to obtain the effective potential for the
classical background as on order parameter with contributions from
quantum fluctuations. By treating $\hat{a} (t)$ and
$\hat{a}^{\dagger} (t)$  as the annihilation and creation
operators, we follow the definition of the coherent state for the
time-independent oscillator \cite{mandel}. The coherent state is
defined as an eigenstate of $\hat{a} (t)$:
\begin{equation}
\hat{a} (t) \vert \alpha, t \rangle = \alpha \vert \alpha, t
\rangle,
\end{equation}
where $\alpha$ is a complex constant. The coherent state can be
treated algebraically by introducing a displacement operator
\begin{equation}
\hat{D}(\alpha) = e^{ - \alpha \hat{a}^{\dagger} (t) + \alpha^*
\hat{a} (t)},
\end{equation}
which is unitary
\begin{equation}
\hat{D} (\alpha) \hat{D}^{\dagger} (\alpha) = \hat{D}^{\dagger}
(\alpha) \hat{D} (\alpha) = \hat{I}.
\end{equation}
Moreover, the displacement operator translates by constants the
annihilation and creation operators through the unitary
transformation
\begin{equation}
\hat{D} (\alpha) \hat{a} (t) \hat{D}^{\dagger} (\alpha) = \hat{a}
(t) + \alpha, \quad \hat{D} (\alpha) \hat{a}^{\dagger} (t)
\hat{D}^{\dagger} (\alpha) = \hat{a} (t) + \alpha^*, \label{un
tran}
\end{equation}
and the inverse unitary transformation
\begin{equation}
\hat{D}^{\dagger} (\alpha) \hat{a} (t) \hat{D} (\alpha) = \hat{a}
(t) - \alpha, \quad \hat{D}^{\dagger} (\alpha) \hat{a}^{\dagger}
(t) \hat{D} (\alpha) = \hat{a} (t) - \alpha^*. \label{inv tran}
\end{equation}
Hence one sees from Eq. (\ref{un tran}) that the coherent state
results from the unitary transformation of the vacuum state
\begin{eqnarray}
\vert \alpha, t \rangle &=& \hat{D}^{\dagger} (\alpha) \vert 0, t
\rangle \nonumber\\ &=& e^{- \frac{\alpha^* \alpha}{2}} \sum_{n =
0}^{\infty} \frac{\alpha^n}{\sqrt{n!}} \vert n, t \rangle.
\label{coh st}
\end{eqnarray}
It has been known since Schr\"{o}dinger that the coherent state
gives rise to a classical field for the time-independent
oscillator \cite{schrodinger}. In our time-dependent case, from
the expectation values of the position and momentum with respect
to the coherent state
\begin{eqnarray}
q_c (t) = \langle \alpha, t \vert \hat{q} \vert \alpha, t \rangle
&=& \hbar \Bigl( \alpha u(t) + \alpha^* u^* (t) \Bigr),
\nonumber\\ p_c (t) = \langle \alpha, t \vert \hat{p} \vert
\alpha, t \rangle &=& \hbar m(t) \Bigl(\alpha \dot{u}(t) +
\alpha^* \dot{u}^* (t)\Bigr),
\end{eqnarray}
one sees that $q_c$ indeed satisfies the classical equation of
motion (\ref{cl sol}), because $u (t)$ has already satisfied Eq.
(\ref{cl sol}) and $\alpha$ is a constant. Besides, $q_c$ is real
and $p_c = m(t) \dot{q}_c$, so one may identify $q_c$ and $p_c$
with the classical position and momentum.

In the above the coherent state has been constructed from the
exact Fock space. There is another method to find the coherent
state based on the minimization of action \cite{cooper}. In
contrast with the LvN approach in which the operators
(\ref{an-cr}) are required to satisfy the LvN equation, one
regards $u$ as a free parameter, works on the $u$-parameter Fock
space and minimizes the Hamiltonian expectation value with respect
to the coherent state. To show the method in detail, take the
Hamiltonian expectation value with respect to the coherent state
\begin{equation}
H_{\rm C} (t) = \langle \alpha, t \vert \hat{H} (t) \vert \alpha,
t \rangle = H_{c} (t) + H_{\rm V} (t),
\end{equation}
which consists of the classical part
\begin{equation}
H_{c} (t) = \frac{1}{2m (t)}p_c^2 + \frac{m(t)}{2}  \omega^2 (t)
q_c^2, \label{cl ham}
\end{equation}
and the vacuum fluctuation part given in Eq. (\ref{num ex}) with
$n = 0$
\begin{equation}
H_{\rm V} (t) = \frac{\hbar^2}{2} m(t) \Bigl[\dot{u}^* (t) \dot{u}
(t) + \omega^2 (t) u^* (t) u(t) \Bigr]. \label{vac ham}
\end{equation}
By writing the complex $u$ in a polar form
\begin{equation}
u(t) = \frac{\xi (t)}{\sqrt{\hbar}} e^{-i \theta (t)},
\label{polar}
\end{equation}
in terms of which Eq. (\ref{boun con}) becomes $\dot{\theta} =
1/(2 m \xi^2)$, and by introducing $p_{\xi} = m (t) \dot{\xi}$,
one obtains the effective Hamiltonian
\begin{equation}
H_{\rm C} (t) = \frac{1}{2m (t)} p_c^2 + \frac{m(t)}{2} \omega^2
(t) q_c^2 + \hbar \Biggl[\frac{1}{2m (t)}p_{\xi}^2 +
\frac{m(t)}{2} \omega^2 (t) \xi^2  + \frac{1}{8 m(t)
\xi^2}\Biggr]. \label{coh ham}
\end{equation}
The last term in the square bracket of Eq. (\ref{coh ham}) has the
same form as the angular momentum for a particle having rotational
symmetry in two dimensions, but its origin is rooted on the
condition (\ref{boun con}) from quantization. Thus the effective
Hamiltonian from the coherent state is equivalent to a
two-dimensional Hamiltonian, which consists of the classical and
quantum fluctuation parts. The variables $q_c$ and $\xi$ are
independent and the Hamilton equations are for $q_c$
\begin{eqnarray}
\frac{dq_c}{dt} &=& \frac{\partial H_{\rm C}}{\partial p_c}  =
\frac{1}{m(t)} p_c, \nonumber\\ \frac{dp_c}{dt} &=& -
\frac{\partial  H_{\rm C}}{\partial q_c} = - m(t) \omega^2 (t)
q_c,
\end{eqnarray}
and for $\xi$
\begin{eqnarray}
\frac{d{\xi}}{dt} &=& \frac{\partial  (H_{\rm C}/\hbar)}{\partial
p_{\xi}} = \frac{1}{m(t)} p_{\xi}, \nonumber\\ \frac{dp_{\xi}}{dt}
&=& - \frac{\partial (H_{\rm C}/\hbar)}{\partial {\xi}} = - m(t)
\omega^2 (t) \xi + \frac{1}{4 m(t) \xi^3}. \label{ham eq}
\end{eqnarray}
It is then easy to show that the Hamilton equations (\ref{ham eq})
equal to the second order equation
\begin{equation}
\ddot{\xi} (t) + \frac{\dot{m} (t)}{m(t)} \dot{\xi} (t) + \omega^2
(t) \xi - \frac{1}{4m^2 (t) \xi^3} = 0, \label{ham eq2}
\end{equation}
and that Eq. (\ref{ham eq2}) is nothing but the equation (\ref{cl
sol}) when $u$ has the form (\ref{polar}) and satisfies the
condition (\ref{boun con}). Hence the minimization of the
effective action gives the identical result as the LvN approach.
Still another method is the mean field approach, in which the
position and momentum are divided into a classical background and
a fluctuation part
\begin{equation}
q = q_c + q_f, \quad p = p_c + p_f.
\end{equation}
Then the total Hamiltonian is composed of three parts: the
classical background and fluctuation parts
\begin{equation}
H (t) = H_{c} (t) + H_{f} (t) + H_{int} (t), \label{cl-fluc}
\end{equation}
where
\begin{equation}
H_f (t) = \frac{1}{2m (t)}p_f^2 + \frac{m(t)}{2} \omega^2 (t)
q_f^2 \label{fluc}
\end{equation}
is the fluctuation Hamiltonian, and
\begin{equation}
H_{int} (t) = \frac{1}{m (t)} p_c p_f + m(t) \omega^2 (t) q_c q_f
\end{equation}
is the interaction Hamiltonian between the classical background
and the fluctuation. And then quantize the fluctuation Hamiltonian
(\ref{fluc}) according to the method in the previous subsection
but keep the classical one unquantized. Since the last two terms
proportional to $q_f$ and $p_f$ have the zero expectation value,
the expectation value of the total Hamiltonian (\ref{cl-fluc})
with respect to the vacuum state of the fluctuation Hamiltonian
(\ref{fluc}) yields exactly the effective Hamiltonian (\ref{coh
ham}). Therefore, it has been shown that the expectation value of
the original Hamiltonian with respect to the coherent state is
equivalent to the sum of the classical part (\ref{cl ham}) and the
vacuum expectation value of the fluctuation part (\ref{fluc}).

\subsection{Thermal State and Density Matrix}

The ensemble of time-dependent oscillators exhibits intrinsically
nonequilibrium behaviors, so it does lose a rigorous physical
meaning to attribute any thermal property to the density operator
(\ref{ham den}). However, the harmonic oscillator problem is
exactly solvable, so even in the time-dependent oscillator case
one may look for a density operator which is quadratic in the
position and momentum, and then fix their variable coefficients to
satisfy the LvN equation \cite{eboli}. On the other hand, in the
LvN approach we can still use the operators (\ref{an-cr}) that
have already satisfied the LvN equation and define the density
operator from them. But there still remains a free parameter to
incorporate the initial thermal equilibrium. We study the physical
meaning of the density operator and see how the initial thermal
equilibrium evolves quantum mechanically .

By noting that $\hat{N} (t)$ satisfies Eq. (\ref{ln eq}), we
define the density operator by
\begin{equation}
\hat{\rho}_{\rm T} (t) = \frac{1}{Z_N}e^{- \beta \hbar \omega_0
(\hat{N} (t) + \frac{1}{2})}, \label{den op}
\end{equation}
where $\beta$ and $\omega_0$ are free parameters and $Z_N$ is the
partition function given by
\begin{equation}
Z_N = \sum_{n = 0}^{\infty} \langle n, t \vert e^{- \beta \hbar
\omega_0 (\hat{N} (t) + \frac{1}{2})} \vert n, t \rangle =
\frac{1}{2\sinh(\frac{\beta \hbar \omega_0}{2})}.
\end{equation}
It has the same form as the standard density operator, the
time-independent annihilation and creation operators now being
replaced by the time-dependent ones (\ref{an-cr}). So Eq.
(\ref{den op}) includes the time-independent case as a special
case by choosing $\beta = 1/(k_B T)$ and $\omega_0$ the oscillator
frequency. In the coordinate representation the density matrix is
given by (see Appendix B)
\begin{eqnarray}
\rho_{\rm T} (q', q, t) &=& \frac{1}{Z_N} \sum_{n = 0}^{\infty}
\Psi_n (q', t) \Psi^*_n (q, t) e^{- \beta \hbar \omega_0 (n +
\frac{1}{2})}\nonumber\\ &=& \Biggl[\frac{\tanh(\frac{\beta \hbar
\omega_0}{2})}{2 \pi \hbar^2 u^* u}\Biggr]^{1/2}  \exp \Biggl[
\frac{i}{4}\frac{m}{\hbar}\frac{d}{dt} \ln (u^* u) (q'^2 - q^2)
\Biggr] \nonumber\\ && \times \exp \Biggl[-\frac{1}{8 \hbar^2 u^*
u} \Biggl\{ (q' + q)^2 \tanh(\frac{\beta \hbar \omega_0}{2}) +
(q'-q)^2 \coth(\frac{\beta \hbar \omega_0}{2}) \Biggr\} \Biggr].
\label{den mat}
\end{eqnarray}
Now the density matrix (\ref{den mat}) can be compared with that
for the time-independent oscillator and the density operator
(\ref{ham den}) for the instantaneous Hamiltonian. For that
purpose we restrict our attention to the particular case, in which
the mass is constant, $m(t) = m_0$, and $\omega^2 (t)$ is positive
(see Sec. IV for the sign changing case of $\omega^2 (t)$) and
slowly changing $|\dot{\omega}(t)/\omega(t)| \ll 1$. In that case
we may look for the solution to Eq. (\ref{cl sol}) of the form
\begin{equation}
u (t) = \frac{1}{\sqrt{2 \hbar m_0 \Omega (t)}} e^{- i \int \Omega
(t)},
\end{equation}
where
\begin{equation}
\Omega^2 (t) = \omega^2 (t) + \frac{3}{4} \frac{\dot{\Omega}^2
(t)}{\Omega^2 (t)} - \frac{1}{2} \frac{\ddot{\Omega}(t)}{\Omega
(t)}. \label{wkb}
\end{equation}
The adiabatic (WKB) solution is obtained by approximating $\Omega
(t) \approx \omega(t)$. Then the density matrix (\ref{den mat})
reduces to the adiabatic one
\begin{eqnarray}
\rho_{\rm A} (q', q, t) &=& \Biggl[\frac{m_0 \omega (t)
\tanh(\frac{\beta \hbar \omega_0}{2})}{\pi \hbar}\Biggr]^{1/2}
\exp \Biggl[-
\frac{i}{4}\frac{m_0}{\hbar}\frac{\dot{\omega}(t)}{\omega (t)}
(q'^2 - q^2) \Biggr] \nonumber\\ && \times \exp \Biggl[-\frac{m_0
\omega (t)}{4 \hbar} \Biggl\{ (q' + q)^2 \tanh(\frac{\beta \hbar
\omega_0}{2}) + (q'-q)^2 \coth(\frac{\beta \hbar \omega_0}{2})
\Biggr\} \Biggr]. \label{den mat2}
\end{eqnarray}
On the other hand, the density operator (\ref{ham den}) for the
instantaneous Hamiltonian has the matrix representation
\begin{eqnarray}
\rho_{\rm H} (q', q, t) &=& \Biggl[\frac{m_0 \omega (t)
\tanh(\frac{\beta \hbar \omega (t)}{2})}{\pi \hbar}\Biggr]^{1/2}
\nonumber\\ && \times \exp \Biggl[-\frac{m_0 \omega (t)}{4 \hbar}
\Biggl\{ (q' + q)^2 \tanh(\frac{\beta \hbar \omega (t)}{2}) +
(q'-q)^2 \coth(\frac{\beta \hbar \omega (t)}{2}) \Biggr\} \Biggr].
\label{den mat3}
\end{eqnarray}

In the case of the time-independent oscillator with $\omega (t) =
\omega_0$, the density matrices (\ref{den mat}) and (\ref{den
mat3}) reduce further to the standard one \cite{kubo}. The
instantaneous density matrix is compared with the adiabatic one by
taking the ratio
\begin{eqnarray}
&&\frac{\rho_{\rm H} (q', q, t)}{\rho_{\rm A} (q', q, t)} =
\Biggl[\frac{\tanh(\frac{\beta \hbar \omega
(t)}{2})}{\tanh(\frac{\beta \hbar \omega_0}{2})} \Biggr]^{1/2}
\exp \Biggl[
\frac{i}{4}\frac{m_0}{\hbar}\frac{\dot{\omega}(t)}{\omega (t)}
(q'^2 - q^2) \Biggr] \nonumber\\&& \times \exp \Biggl[-\frac{m_0
\omega (t)}{4 \hbar} \Biggl\{ (q' + q)^2
\frac{\sinh\Bigl(\frac{\beta \hbar}{2} (\omega (t) - \omega_0)
\Bigr)}{\cosh(\frac{\beta \hbar \omega (t)}{2})\cosh(\frac{\beta
\hbar \omega_0}{2})} - (q'-q)^2 \frac{\sinh\Bigl(\frac{\beta
\hbar}{2} (\omega (t) - \omega_0) \Bigr)}{\sinh(\frac{\beta \hbar
\omega (t)}{2})\sinh(\frac{\beta \hbar \omega_0}{2})} \Biggr\}
\Biggr].\label{ratio}
\end{eqnarray}
The second factor in Eq. (\ref{ratio}) gives rise to a small phase
factor because $\omega (t)$ is slowly varying. As far as $\omega
(t)$ remains close to $\omega_0$, the instantaneous density matrix
(\ref{den mat3}) is close to the adiabatic one (\ref{den mat2}).
Otherwise, the exact nonequilibrium evolution (\ref{den mat2}) is
far away from the quasi-equilibrium one described by (\ref{den
mat3}). This implies a significant deviation of the nonequilibrium
evolution from the equilibrium one as $\omega(t)$ differs from
$\omega_0$ by a large amount.

Being mostly interested in phase transitions, we assume the system
to start from an initial thermal equilibrium at early times. This
requires that the oscillator have a constant frequency $\omega_0$
at early times and $\beta$ be fixed to the inverse temperature. As
in the case of coherent state, the evolution of the initial
thermal state can be found the effective Hamiltonian from this
thermal state. From the expectation values (see the appendix of
Ref. \cite{kim2})
\begin{eqnarray}
\langle \hat{q}^{2} \rangle_{\rm T} &=& {\rm Tr}
\Bigl[\hat{\rho}_{\rm T} (t) \hat{q}^{2} \Bigr] = \hbar^2 u^*(t)
u(t)\coth(\frac{\beta \hbar \omega_0 }{2}), \nonumber\\ \langle
\hat{p}^{2} \rangle_{\rm T} &=& {\rm Tr} \Bigl[\hat{\rho}_{\rm T}
(t) \hat{p}^{2} \Bigr] = \hbar^2 m^2(t) \dot{u}^*(t)
\dot{u}(t)\coth(\frac{\beta \hbar \omega_0 }{2}),
\end{eqnarray}
one obtains the effective Hamiltonian from the thermal state
\begin{equation}
H_{\rm T}(t) = {\rm Tr}\Bigl[\hat{\rho}_{\rm T} (t) \hat{H} (t)
\Bigr] = \frac{\hbar^2}{2} m(t) \coth(\frac{\beta \hbar \omega_0
}{2}) \Bigl[ \dot{u}^* (t) \dot{u} (t) + \omega^2 (t) u^*(t) u(t)
\Bigr].
\end{equation}
Once again by using the complex parameter $u$ in the polar form
(\ref{polar}) and by introducing $p_{\xi} = m (t) \dot{\zeta}$,
one may rewrite the effective Hamiltonian as
\begin{equation}
H_{\rm T} (t) = \hbar \coth(\frac{\beta \hbar \omega_0 }{2})
\Biggl[\frac{1}{2m (t)}p_{\xi}^2 + \frac{m(t)}{2} \omega^2 (t)
\zeta^2 + \frac{1}{8 m(t) \xi^2}\Biggr] = \coth(\frac{\beta \hbar
\omega_0 }{2}) H_{\rm V} (t). \label{ther ham}
\end{equation}
Since $\hbar \coth(\beta \hbar \omega_0 / 2)$ is constant, $H_{\rm
T} (t)$ has the same Hamilton equations (\ref{ham eq2}) as $H_{\rm
V} (t)$. This is identical to the classical equation of motion
(\ref{cl sol}) together with the boundary condition (\ref{boun
con}).

\subsection{Coherent-Thermal State}

A more general density operator that is at most quadratic in the
position and momentum was introduced in Ref. \cite{kim2}. It has
the form
\begin{equation}
\hat{\rho}_{\rm C.T} (t) = \frac{1}{Z_{\rm C.T} (t)} \exp \Bigl[-
\beta \Bigl\{\hbar \omega_0 \hat{a}^{\dagger} (t) \hat{a}(t) +
\delta \hat{a}^{\dagger} (t) + \delta^* \hat{a} (t) +
\epsilon_0\Bigr\} \Bigr], \label{c-t den}
\end{equation}
where $Z_{\rm C.T}$ is a partition function. In fact, the density
operator of Eq. (\ref{c-t den}) can be transformed into that of
Eq. (\ref{den op}) by the unitary transformation
\begin{equation}
\hat{D}^{\dagger} (\alpha) \hat{\rho}_{\rm C} (t) \hat{D} (\alpha)
= \hat{\rho} (t),
\end{equation}
where $\hat{D} (\alpha)$ with $\alpha = \delta/(\hbar \omega_0)$
and $\epsilon_0 = (\hbar \omega_0/2) + (|\delta|^2/\hbar
\omega_0)$. From the expectation values
\begin{eqnarray}
\langle \hat{q}^{2} \rangle_{{\rm C.T}} &=& {\rm Tr}
\Bigl[\hat{\rho}_{\rm C.T} (t) \hat{q}^{2} \Bigr] =  q_c^2 +
\hbar^2 u^*(t) u(t)\coth(\frac{\beta \hbar \omega_0 }{2}),
\nonumber\\ \langle \hat{p}^{2} \rangle_{{\rm C.T}} &=& {\rm Tr}
\Bigl[\hat{\rho}_{\rm C.T} (t) \hat{p}^{2} \Bigr] =  p_c^2 +
\hbar^2 m^2 (t) \dot{u}^*(t) \dot{u}(t)\coth(\frac{\beta \hbar
\omega_0 }{2}),
\end{eqnarray}
follows the effective Hamiltonian
\begin{equation}
H_{\rm C.T} (t) = H_c (t) + H_{\rm T} (t), \label{c-t ham}
\end{equation}
where $H_c$ is the classical Hamiltonian. The effective
Hamiltonian (\ref{c-t ham}) has almost the same form as Eq.
(\ref{coh ham}) except for the overall factor $\coth(\beta \hbar
\omega_0/2)$, hence describes the same Hamilton equations
(\ref{ham eq}).

In summary, what we have shown in this section is that the LvN
approach provides us with the exact Fock space and various quantum
states for time-dependent oscillators. The only auxiliary field
necessary for that purpose is a complex solution $u$ to the
classical equation of motion (\ref{cl sol}) that satisfies the
wronskian condition (\ref{boun con}) from quantization. It has
further been shown that the LvN approach is equivalent to the
minimization principle for the effective action and to the mean
field method. However, the advantage of the LvN approach lies in
the manifest correspondence with quantum mechanics and quantum
many-particle system and in the readiness to apply their standard
techniques. For instance, the density matrix (\ref{den mat}) has
been found, which has many features similar with the standard one.

\section{Inverted Harmonic Oscillator}

As a toy model for the second order phase transition, let us
consider the time-dependent harmonic oscillator
\begin{equation}
\hat{H} = \frac{1}{2} \hat{p}^2 + \frac{1}{2} \omega^2 (t)
\hat{q}^2,
\end{equation}
where $\omega^2 (t)$ has the asymptotic value $\omega^2_i (> 0)$
far before and $- \omega^2_f (<0)$ far after the quench. The
conspicuous point of the model is the sign change of $\omega^2
(t)$. At earlier times before the quench the oscillator executes a
stable motion about $q = 0$, the global minimum, but after the
quench the potential is inverted and $q = 0$ becomes an unstable
configuration.

At earlier times far before the phase transition, the complex
solution to Eq. (\ref{cl sol}) satisfying Eq. (\ref{boun con}) is
given by
\begin{equation}
u_i(t) = \frac{e^{-i \omega_i t}}{\sqrt{2 \hbar \omega_i}}.
\label{osc asym}
\end{equation}
According to Eq. (\ref{an-cr}), the Fock space is now constructed
from the annihilation and creation operators
\begin{eqnarray}
\hat{a} (t) &=& \frac{e^{i \omega_i t}}{\sqrt{2 \hbar \omega_i}}
\Bigl(i \hat{p} + \omega_i \hat{q} \Bigr) = e^{i \omega_i t}
\hat{a}_0, \nonumber\\ \hat{a}^{\dagger} (t) &=&  \frac{e^{- i
\omega_i t}}{\sqrt{2 \hbar \omega_i}} \Bigl(-i \hat{p} + \omega_i
\hat{q} \Bigr) = e^{- i \omega_i t} \hat{a}^{\dagger}_0.
\end{eqnarray}
Note that $\hat{a} (t)$ and $\hat{a}^{\dagger} (t)$ differ from
the standard ones $\hat{a}_0$ and $\hat{a}_0^{\dagger}$ only by
phase factors. Though the Hamiltonian has the standard
representation
\begin{equation}
\hat{H}_i = \hbar \omega_i \Bigl( \hat{a}^{\dagger} (t) \hat{a}
(t) + \frac{1}{2} \Bigr) =  \hbar \omega_i \Bigl(
\hat{a}^{\dagger}_0 \hat{a}_0 + \frac{1}{2} \Bigr),
\end{equation}
the phase factors are necessary for $\hat{a} (t)$ and
$\hat{a}^{\dagger} (t)$ to satisfy the LvN equation. Hence the
vacuum expectation value is given by the well-known result
\begin{equation}
H_{\rm V} = \frac{1}{2} \hbar \omega_i,
\end{equation}
and the coherent state (\ref{coh st}) yields
\begin{equation}
H_{\rm C} = \frac{1}{2} p_c^2 + \frac{1}{2} \omega_i^2 q_c^2 +
\frac{1}{2} \hbar \omega_i.
\end{equation}
Now the density operator (\ref{den op}) reduces to the standard
one
\begin{equation}
\hat{\rho}_{i, {\rm T}} = \frac{1}{Z_N} e^{- \beta \hbar \omega_i
(\hat{a}_0^{\dagger} \hat{a}_0 + \frac{1}{2})},
\end{equation}
after identifying $\omega_0 = \omega_i$ and $\beta = 1/(k_B T)$,
and leads to the Hamiltonian expectation value
\begin{equation}
H_{i, {\rm T}} = \frac{1}{2} \hbar \omega_i \coth(\frac{\beta
\hbar \omega_i}{2}).
\end{equation}

On the other hand, at later times far after the quench, the
solution to Eq. (\ref{cl sol}) is given by
\begin{eqnarray}
u_f (t) &=& \frac{1}{\sqrt{2 \hbar}} \Bigl[ C_1(\omega_i,
\omega_f) \cosh(\omega_f t) - i C_2(\omega_i, \omega_f )
\sinh(\omega_f t) \Bigr] \nonumber\\ &=& \frac{1}{2 \sqrt{2
\hbar}} \Bigl[ \Bigl( C_1(\omega_i, \omega_f) - i C_2(\omega_i,
\omega_f) \Bigr) e^{\omega_f t} +  \Bigl( C_1(\omega_i, \omega_f)
+ i  C_2(\omega_i, \omega_f) \Bigr) e^{- \omega_f t} \Bigr],
\label{osc asym2}
\end{eqnarray}
where $C_j, j = 1, 2$ depend on the intermediate process toward
the final state. Remarkably, the vacuum and thermal expectation
values vanish:
\begin{equation}
H_{f, {\rm V}} = H_{f, {\rm T}} = 0,
\end{equation}
since the kinetic and potential energies contribute equally.
Whereas the expectation value with respect to the coherent state
(\ref{coh st}) and the general density operator (\ref{c-t den})
takes the form
\begin{equation}
H_{f, {\rm C}} = H_{f, {\rm C.T}} = \frac{1}{2} p_c^2 -
\frac{1}{2} \omega_f^2 q_c^2.
\end{equation}
This implies physically that as the system undergoes the phase
transition out of equilibrium, quantum effects vanish and it
becomes classical. Large uncertainty has been suggested as one of
the criteria on the classicality \cite{guth}
\begin{equation}
(\Delta q) (\Delta p) = \frac{\hbar}{2} \Bigl[ \omega_f C_1^2
(\omega_i, \omega_f) \cosh^2 (\omega_f t)  + \omega_f C_1^2
(\omega_i, \omega_f) \sinh^2 (\omega_f t)\Bigr].
\end{equation}
Though the oscillator starts with the minimum uncertainty given by
Eq. (\ref{osc asym}), its uncertainty increases exponentially and
it becomes eventually classical after the completion of quench.
When Eq. (\ref{osc asym2}) is substituted into Eq. (\ref{den
mat}), the density matrix depends on the intermediate process and
spreads as $\sqrt{u_f^* (t) u_f (t)}$. Therefore the quintessence
of second order phase transitions lies in the whole process how
systems evolve out of equilibrium from their initial equilibrium.

To show explicitly how the nonequilibrium dynamics depends on the
intermediate processes, we consider two exactly solvable models.
The first model, which is the zero mode of the free scalar model
in Sec. VI, is an oscillator describing a finite smooth quench
with the mass
\begin{equation}
m^2 (t) = m_1^2 - m_0^2 \tanh \Bigl(\frac{t}{\tau} \Bigr).
\end{equation}
The mass has $m_i^2 = m_0^2 + m_1^2$ at earlier times $(t = -
\infty)$ and has $- m_f^2 = - m_0^2 + m_1^2 < 0$ at later times
$(t = \infty)$. $\tau$ measures the quench rate, {\it i.e.} the
rate of change of mass. The instantaneous quench corresponds to
the $(\tau = 0)$-limit. The solution to Eq. (\ref{cl sol}) is
found
\begin{equation}
u (t) = \frac{e^{- m_i t}}{\sqrt{2 \hbar m_i}}  {}_2F_1
\Bigl(-\frac{\tau}{2} (i m_i - m_f), -\frac{\tau}{2} (i m_i +
m_f); 1 - i \tau m_i; - e^{2t/\tau}\Bigr). \label{mod sol}
\end{equation}
At earlier times the solution (\ref{mod sol}) has the correct
asymptotic form
\begin{equation}
u_i (t) = \frac{e^{-i m_i t}}{\sqrt{2 \hbar m_i}}. \label{mod
asym}
\end{equation}
On the other hand, at later times the asymptotic form of Eq.
(\ref{mod sol}) becomes \cite{abramowitz}
\begin{eqnarray}
u_f (t) &=&  \Biggl[\frac{1}{\sqrt{2 \hbar m_i}} \frac{(-1) \Gamma
(1-i m_i\tau) \Gamma (m_f\tau)}{\frac{\tau}{2}(i m_i-
m_f)\Gamma^2(- \frac{\tau}{2}(i m_i - m_f)} \Biggr] e^{m_f t}
\nonumber\\ && + \Biggl[ \frac{1}{\sqrt{2 \hbar m_i}} \frac{(-1)
\Gamma (1-i m_i\tau) \Gamma (m_f\tau)}{\frac{\tau}{2}(i m_i+
m_f)\Gamma^2(- \frac{\tau}{2}(i m_i + m_f)} \Biggr] e^{- m_f t}.
\label{mod asym2}
\end{eqnarray}
Two points are observed: the initial solution branches into an
unstable growing and a decaying mode as expected and the
coefficients $C_1, C_2$ of Eq. (\ref{osc asym2}) depend on the the
mass parameters $m_i, m_f$ and the quench rate $\tau$. In other
words, the final asymptotic state of nonequilibrium evolution
depends on the intermediate process.

The next model describes various quench processes and exhibits how
the stable mode far before the quench branches into the unstable
growing and decaying modes during the quench. Without loss of
generality the nonequilibrium phase transition is assumed to take
place through the time-dependent frequency (mass) squared
\begin{eqnarray}
\omega^2 (t)  = \cases{ \omega_i^2, & $t_i > t$, \cr \omega_i^2
\Biggl(\frac{t_0 - t}{t_0 - t_i}\Biggr)^{(2l_1+1)/(2l_2+1)}, &
$t_0 > t > t_i$, \cr - \omega_i^2 \Biggl(\frac{t - t_0}{t_0 -
t_i}\Biggr)^{(2l_1+1)/(2l_2+1)}, & $t_f > t > t_0$, \cr -
\omega_f^2 \equiv - \omega_i^2 \Biggl(\frac{t_f - t_0}{t_0 -
t_i}\Biggr)^{(2l_1+1)/(2l_2+1)}, & $t > t_f$, \cr} \label{gen
freq}
\end{eqnarray}
where $l_1$ and $l_2$ are non-negative integers. Here $t_0 - t_i$
adjusts the rate of and $t_f - t_0$ the duration of the quench.
The particular case of $l_1 = l_2 = 0$ is used as the finite
linear quench model \cite{bowick,rivers}. Before the time $t_0$,
the system maintains the symmetry about $q = 0$, the minimum of
the potential. But as time goes on after $t_0$, $q = 0$ remains no
longer the true minimum of the system and the symmetry is broken.
The particular form of the power-law in Eq. (\ref{gen freq}) is
chosen to allow an analytical continuation of $\omega^2(t)$ for
changing the sign and to make its derivatives also continuous.
Before $t_i > t$, the solution is given by Eq. (\ref{osc asym}).
During $t_0 > t
> t_i$, the solution is given by a linear superposition of Hankel
functions
\begin{equation}
u (t) = D_1 z^{\nu} H^{(2)}_{\nu} (z) + D_2 z^{\nu} H^{(1)}_{\nu}
(z), \label{gen sol}
\end{equation}
where
\begin{equation}
z = 2 \nu \omega_i (t_0 - t_i) \Biggl(\frac{t_0 - t}{t_0 -
t_i}\Biggr)^{1/2\nu}, \quad \nu = \frac{2l_2+1}{2l_1 + 4 l_2 + 3}.
\end{equation}
Here $H^{(2)}_{\nu}$ and $H^{(1)}_{\nu}$ are positive and negative
frequency solutions, respectively. The constants $D_1$ and $D_2$
are determined by continuity of $u(t)$ and $ \dot{u} (t)$ across
$t_i$:
\begin{eqnarray}
D_1 &=& \frac{e^{-i \omega_i t_i}}{\sqrt{2 \hbar \omega_i}}
\frac{\pi}{4 z_i^{\nu}} \Biggl[- i z_i \frac{d}{dz_i}
H^{(1)}_{\nu} (z_i) - \Bigl(\omega_i - i \frac{1}{2 (t_0 - t_i)}
\Bigr) H^{(2)}_{\nu} (z_i) \Biggr], \nonumber\\ D_2 &=&
\frac{e^{-i \omega_i t_i}}{\sqrt{2 \hbar \omega_i}} \frac{\pi}{4
z_i^{\nu}} \Biggl[ i z_i \frac{d}{dz_i} H^{(2)}_{\nu} (z_i) +
\Bigl(\omega_i - i \frac{1}{2 (t_0 - t_i)} \Bigr) H^{(1)}_{\nu}
(z_i) \Biggr], \label{gen coef}
\end{eqnarray}
where
\begin{equation}
z_i = 2 \nu \omega_i (t_0 - t_i).
\end{equation}
Beyond the quench time $t_0$, it is necessary to do carefully the
analytic continuation and to take a suitable Riemann sheet so that
$\omega^2 (t)$ and its derivatives are to be continuous from
$\omega^2> 0$ to $\omega^2 < 0$ across $t_0$. The analytic
continuation of the solution (\ref{gen sol}) yields
\begin{eqnarray}
u (t) &=& \frac{D_1}{2} \tilde{z}^{\nu} \Biggl[e^{i 3\pi
(l_2+\frac{1}{2})} H^{(2)}_{\nu} (i \tilde{z}) + e^{i\pi
(l_2+\frac{1}{2})} H^{(2)}_{\nu} (-i \tilde{z})\Biggr] \nonumber\\
&& + \frac{D_2}{2} \tilde{z}^{\nu} \Biggl[e^{i 3\pi
(l_2+\frac{1}{2})} H^{(1)}_{\nu} (i \tilde{z}) + e^{i\pi
(l_2+\frac{1}{2})} H^{(1)}_{\nu} (-i \tilde{z})\Biggr], \label{gen
sol2}
\end{eqnarray}
where
\begin{equation}
\tilde{z} = 2 \nu \omega_i (t_0 - t_i) \Biggl(\frac{t - t_0}{t_0 -
t_i}\Biggr)^{1/2\nu} .
\end{equation}
At later times $(\tilde{z} \gg 1)$ during the quench, the solution
(\ref{gen sol2}) has the asymptotic form \cite{gradshteyn}
\begin{equation}
u_f (t) = \sqrt{\frac{1}{2 \pi}} \tilde{z}^{\nu - \frac{1}{2}}
e^{\tilde{z}} \Biggl[ D_1 e^{i\pi \bigl(3l_2+ \frac{\nu}{2} +
\frac{3}{2} \bigr)} + D_2 e^{i\pi \bigl(l_2 - \frac{\nu}{2} -
\frac{1}{2} \bigr)} \Biggr]. \label{gen asym2}
\end{equation}
Thus the stable mode of the oscillating solution (\ref{osc asym})
branches into the growing mode (\ref{gen asym2}) which dominates
during various quench processes and into the decaying mode which
contributes negligibly to the correlation functions. The
asymptotic solution (\ref{gen asym2}) also depends on the
intermediate processes through $t_0, t_i, t_f$ and $l_1, l_2$.

\section{Time-Dependent Anharmonic Oscillator}

In this section we extend the formalism developed in Sec. III to
the time-dependent anharmonic oscillator with the Hamiltonian
\begin{equation}
H(t) = \frac{p^2}{2 m(t)} + m(t) V(q), \label{an osc}
\end{equation}
where
\begin{equation}
V (q) = \frac{\lambda_{2n} (t) }{(2n)!}q^{2n}.
\end{equation}
Though the potential of a power law is assumed, the formalism can
readily be generalized to any polynomial and analytic potential.
In the time-independent case $(\lambda_{2n} = {\rm constant})$,
the variational perturbation method has been introduced as one of
the powerful methods to find the Hilbert space \cite{chang}. The
vacuum state in this approach is the Gaussian wave functional that
minimizes the effective action. The excited states are then
obtained from the vacuum state just as number states of a harmonic
oscillator are obtained from the ground state. Though these states
can be calculated explicitly in terms a complex solution to the
classical equation for motion, they are in fact equivalent to
those from the mean-field method.

However, there have been some attempts to go beyond the Gaussian
state. In Ref. \cite{kim8} a scheme was proposed to find the
operators, which generalize the annihilation and creation
operators and satisfy the LvN equation (\ref{ln eq}), to all the
orders of coupling constant in the time-independent case and to
the first order in the time-dependent case. In particular, the
generalized annihilation and creation operators for the
time-independent oscillator with a quartic potential satisfy a
q-deformed algebra rather than the standard commutation relation
\cite{kim9}, from which follows an algebraic construction of
excited states and energy spectra beyond the variational Gaussian
approximation. It would be interesting to find such an algebraic
structure for interacting quantum fields, which may shed some
light on the nonperturbative method beyond the mean-field method.
Also it would be interesting to compare this scheme with other
nonperturbative methods in the time-independent case such as the
perturbative expansion method around the Gaussian effective action
\cite{yee} and the time-dependent variational method
\cite{cheetham}. But we shall not pursue further this issue in
this paper.

\subsection{Fock Space}

In the case of time-dependent anharmonic oscillators, the LvN
approach searches for the annihilation and creation operators that
are still linear in the position and momentum and satisfy the LvN
equation (\ref{ln eq}):
\begin{eqnarray}
\hat{A} (t) &=& i \Bigl( v^* (t) \hat{p} - m (t) \dot{v}^* (t)
\hat{q}\Bigr), \nonumber\\ \hat{A}^{\dagger} (t) &=& - i \Bigl( v
(t) \hat{p} - m (t) \dot{v} (t) \hat{q}\Bigr). \label{cr-an 2}
\end{eqnarray}
One then requires them to satisfy the LvN equation (\ref{ln eq}),
leading to the equation
\begin{equation}
\ddot{v} (t) \hat{q} + \frac{\dot{m}(t)}{m(t)} \dot{v} (t) \hat{q}
+ v(t) \frac{\delta V(\hat{q})}{\delta \hat{q}} = 0,\label{cl eq2}
\end{equation}
and further differentiates functionally with respect to $\hat{q}$
and takes the vacuum expectation value of the resultant equation
\begin{equation}
\ddot{v} (t) + \frac{\dot{m}(t)}{m(t)} \dot{v} (t) +  \langle 0, t
\vert \frac{\delta^2 V(\hat{q})}{\delta \hat{q}^2} \vert 0, t
\rangle v(t) = 0. \label{cl eq3}
\end{equation}
Here the vacuum state is annihilated by $\hat{A} (t)$
\begin{equation}
\hat{A} (t) \vert 0, t \rangle = 0.
\end{equation}
One makes $\hat{A} (t)$ and $\hat{A}^{\dagger} (t)$ the
annihilation and creation operators, respectively, by imposing the
standard commutation relation for all times
\begin{equation}
[ \hat{A} (t), \hat{A}^{\dagger} (t)] = 1. \label{an com}
\end{equation}
Equation (\ref{an com}) is equivalent to the wronskian condition
\begin{equation}
\hbar m(t) \Bigl(\dot{v}^* (t) v(t) - \dot{v} (t) v^* (t) \Bigr) =
i.
\end{equation}
The Fock space consists of the number state obtained by applying
$\hat{A}^{\dagger} (t)$ $n$-times to the vacuum state
\begin{equation}
\vert n, t \rangle = \frac{\Bigl(\hat{A}^{\dagger} (t)
\Bigr)^n}{\sqrt{n!}} \vert 0, t \rangle.
\end{equation}
These number states are excited states and have the coordinate
representation (\ref{har wav}) now with $u(t)$ replaced by $v(t)$.

>From the position and momentum operators expressed in terms of the
annihilation and creation operators
\begin{eqnarray}
\hat{q} = \hbar \Bigl( v(t) \hat{A} (t) + v^* (t)
\hat{A}^{\dagger} (t) \Bigr), \nonumber\\ \hat{p} = \hbar m(t)
\Bigl( \dot{v}(t) \hat{A} (t) + \dot{v}^* (t) \hat{A}^{\dagger}
(t) \Bigr),
\end{eqnarray}
follow the vacuum expectation values
\begin{eqnarray}
\langle 0, t \vert \hat{q}^{2n} \vert 0, t \rangle =
\frac{(2n)!}{2^n n!} \Bigl[\hbar^2 v^*(t) v(t)\Bigr]^{n},
\nonumber\\ \langle 0, t \vert \hat{p}^{2} \vert 0, t \rangle =
\hbar^2 m^2(t) \dot{v}^*(t) \dot{v}(t).
\end{eqnarray}
Then the classical equation of motion (\ref{cl eq3}) becomes
\begin{equation}
\ddot{v} (t) + \frac{\dot{m}(t)}{m(t)} \dot{v} (t) +
\frac{\lambda_{2n} (t)}{2^{n-1} (n-1)!} \Bigl[ \hbar^2 v^*(t) v(t)
\Bigr]^{n-1} v(t) = 0. \label{cl eq4}
\end{equation}
There is another method to derive Eq. (\ref{cl eq4}). By using the
Wick-ordering
\begin{equation}
\hat{q}^{2n} = \sum_{k = 0}^{n} \frac{(2n)! \hbar^{2n}}{2^k k! (2n
- 2k)!} \Bigl[ v^* (t) v(t)]^{k} : [ v(t) \hat{A} (t) + v^* (t)
\hat{A}^{\dagger} (t) \Bigr]^{2(n-k)}:, \label{q-2n}
\end{equation}
one obtains the Hamiltonian truncated at the quadratic order of
$\hat{A}(t)$ and $\hat{A}(t)$
\begin{eqnarray}
\hat{H}_{\rm G} &=& \frac{1}{2}\hbar^2 m(t) \Bigl[(\dot{v}(t)
\hat{A} (t))^2 + 2 \dot{v}^*(t) \dot{v}(t) \hat{A}^{\dagger} (t)
\hat{A} (t) + (\dot{v}^*(t) \hat{A}^{\dagger} (t))^2 \Bigr]
\nonumber\\ &+& m(t) \frac{\hbar^{2n}\lambda_{2n}(t)}{2^n (n-1)!}
[v^* (t) v(t)]^{n-1} \Bigl[(v(t) \hat{A} (t))^2 + 2 v^*(t) v(t)
\hat{A}^{\dagger} (t) \hat{A} (t) + (v^*(t) \hat{A}^{\dagger}
(t))^2 \Bigr], \label{trun ham}
\end{eqnarray}
where purely $c$-number terms are neglected. One then requires
$\hat{A}^{\dagger} (t)$ and $\hat{A}(t)$ to satisfy the LvN
equation (\ref{ln eq}) for the truncated Hamiltonian $\hat{H}_{\rm
G} (t)$. Now the LvN equations for $\hat{A}^{\dagger} (t)$ and
$\hat{A} (t)$ lead exactly to the equation of motion (\ref{cl
eq4}).

\subsection{Effective Hamiltonians}

Still another method to derive the equations of motion for $v$ and
$q_c$ is the minimization principle for the effective action. For
that purpose we consider a complex $v(t)$-parameter family of the
Fock spaces constructed by the annihilation and creation operators
(\ref{cr-an 2}) and take the Hamiltonian expectation value with
respect to various quantum states such as the vacuum, coherent,
thermal and coherent-thermal states. We do not require $\hat{A}
(t)$ and $\hat{A}^{\dagger} (t)$ to satisfy the LvN equation a
priori, but minimize the action to determine the equation of
motion for $v(t)$.

The first effective Hamiltonian is the vacuum expectation value
\begin{equation}
H_{\rm V} (t) = \langle 0, t \vert \hat{H} (t) \vert 0, t \rangle
= \frac{\hbar^2}{2} m(t) \dot{v}^* \dot{v} +
m(t)\frac{\lambda_{2n} (t) }{2^n n!} \Bigl[ \hbar^2 v^*(t) v(t)
\Bigr]^n. \label{an vac ham}
\end{equation}
The next state under consideration is the coherent state, which is
obtained by applying a displacement operator to the vacuum state,
\begin{equation}
\vert \alpha, t \rangle = \hat{D}^{\dagger} (\alpha) \vert 0, t
\rangle = e^{\alpha \hat{A}^{\dagger} (t) - \alpha^* \hat{A} (t)}
\vert 0, t \rangle.
\end{equation}
Then the coherent state expectation value leads to the effective
Hamiltonian
\begin{equation}
H_{\rm C} (t) = \langle \alpha, t \vert \hat{H} (t) \vert \alpha,
t \rangle,
\end{equation}
which, with the aid of Eq. (\ref{q-2n}), is decomposed into
\begin{equation}
H_{\rm C} (t) = H_c (t) + H_{q} (t). \label{an coh ham}
\end{equation}
Here $H_c (t)$ is the classical part
\begin{equation}
H_c(t) = \frac{p^2_c}{2 m(t)} + m(t) \frac{\lambda_{2n} (t)
}{(2n)!}q^{2n}_c,
\end{equation}
and $H_q (t)$ denotes all the quantum contributions including the
vacuum expectation value
\begin{equation}
H_q (t) = \frac{\hbar^2}{2} m(t) \dot{v}^* \dot{v} + m(t) \sum_{k
= 1}^{n} \frac{\lambda_{2n} (t) }{2^k k! (2n- 2k)!} \Bigl[\hbar^2
v^*(t) v(t) \Bigr]^k q^{2(n-k)}_c, \label{coh-q}
\end{equation}
where
\begin{equation}
q_c = \langle \alpha, t \vert \hat{q} \vert \alpha, t \rangle,
\quad p_c = \langle \alpha, t \vert \hat{p} \vert \alpha, t
\rangle.
\end{equation}
The final state is the thermal state defined by the density
operator
\begin{equation}
\hat{\rho}_{\rm T} = \frac{1}{Z_N} e^{- \beta \omega_0
(\hat{A}^{\dagger} (t) \hat{A} (t) + \frac{1}{2})}, \label{an den
op}
\end{equation}
with $Z_N$ being the partition function. The density operator
(\ref{an den op}) leads to the effective Hamiltonian
\begin{equation}
H_{\rm T} (t) = {\rm Tr} \Bigl[\hat{\rho}_{\rm T} \hat{H} (t)
\Bigr] = \frac{\hbar^2}{2} m(t) \dot{v}^* \dot{v} + m(t)
\frac{\lambda_{2n} (t) }{2^n n!} \langle \hat{q}^2 \rangle_{\rm T}
^n, \label{an ther ham}
\end{equation}
where
\begin{equation}
\langle \hat{q}^2 \rangle_{\rm T} = \hbar^2 v^* v
\coth(\frac{\beta \omega_0}{2}).
\end{equation}
Likewise, the density operator of the form (\ref{c-t den}) with
$\hat{a}^{\dagger} (t)$ and $\hat{a} (t)$ replaced by
$\hat{A}^{\dagger} (t)$ and $\hat{A} (t)$ leads to the effective
Hamiltonian
\begin{equation}
H_{\rm C.T}(t) =  H_c (t) + \frac{\hbar^2}{2} m(t)
\coth(\frac{\beta \hbar \omega_0}{2}) \dot{v}^* \dot{v} + m(t)
\sum_{k = 1}^{n} \frac{\lambda_{2n} (t) }{2^k k! (2n- 2k)!}
\langle \hat{q}^2 \rangle_{\rm T}^k q^{2(n-k)}_c. \label{an c-t
ham}
\end{equation}
Note that Eqs. (\ref{an vac ham}) and (\ref{an coh ham}) are also
obtained from Eqs. (\ref{an ther ham}) and (\ref{an c-t ham}),
respectively, by replacing $\hbar^2 v^* v \coth(\beta \omega_0/2)$
with $\hbar^2 v^* v$ or taking the zero-temperature limit $(\beta
\rightarrow \infty)$.

We now study the dynamics of the effective Hamiltonians. We mainly
focus on the effective Hamiltonian (\ref{an c-t ham}) since Eq.
(\ref{an coh ham}) is the limiting case of Eq. (\ref{an c-t ham})
when $\beta \rightarrow \infty$, {\it i.e.}, $T \rightarrow 0$,
and Eq. (\ref{an vac ham}) is the limiting case of Eq. (\ref{an
coh ham}) when $q_c = p_c= 0$. The independent variables of the
Hamiltonian (\ref{an c-t ham}) are $(q_c, p_c)$, $(v, p_v = m
\dot{v}^*)$ and $(v^*, p^*_{v} = m \dot{v})$. So we obtain the
equation of motion for $q_c$
\begin{equation}
\ddot{q}_c + \frac{\dot{m}}{m}\dot{q}_c + \frac{\lambda_{2n} (t)
}{(2n-1)!}q^{2n -1}_c + \sum_{k = 1}^{n}
\frac{\lambda_{2n}(t)}{2^k k! (2n- 2k -1)!} \langle \hat{q}^2
\rangle_{\rm T}^k q^{2n - 2k -1}_c = 0. \label{an osc eq}
\end{equation}
The equation of motion for $v^*$ is given by
\begin{equation}
\ddot{v} + \frac{\dot{m}}{m} \dot{v} + \sum_{k = 1}^{n}
\frac{\lambda_{2n}(t)}{2^{k-1} (k-1)! (2n- 2k)!} q^{2n - 2k}_c
\langle \hat{q}^2 \rangle_{\rm T.}^{k-1} v = 0, \label{an osc eq2}
\end{equation}
and the complex conjugate of Eq. (\ref{an osc eq2}) is for $v$.
The equations of motion from the effective Hamiltonian (\ref{an
coh ham}) is the limiting case of Eqs. (\ref{an osc eq}) and
(\ref{an osc eq2}) when $\langle \hat{q}^2 \rangle_{({\rm T})} =
\hbar^2 v^* v$:
\begin{eqnarray}
\ddot{q}_c + \frac{\dot{m}}{m}\dot{q}_c + \frac{\lambda_{2n} (t)
}{(2n-1)!}q^{2n -1}_c + \sum_{k = 1}^{n}
\frac{\lambda_{2n}(t)}{2^k k! (2n- 2k -1)!} (\hbar^2 v^* v)^k
q^{2n - 2k -1}_c = 0, \nonumber\\ \ddot{v} + \frac{\dot{m}}{m}
\dot{v} + \sum_{k = 1}^{n} \frac{\lambda_{2n}(t)}{2^{k-1} (k-1)!
(2n- 2k)!} q^{2n - 2k}_c (\hbar^2 v^* v)^{k-1} v = 0. \label{an
osc eq3}
\end{eqnarray}
And the limiting case $q_c = 0$ of Eq. (\ref{an osc eq3}) is the
equation for the effective Hamiltonian (\ref{an vac ham})
\begin{equation}
\ddot{v} + \frac{\dot{m}}{m} \dot{v} +
\frac{\lambda_{2n}(t)}{2^{n-1} (n-1)!} (\hbar^2 v^* v)^{n-1} v =
0. \label{an osc eq4}
\end{equation}
Note that Eq. (\ref{an osc eq4}) is identical to Eq. (\ref{cl
eq4}) from the LvN approach.

Or, by writing $v$ in the polar form
\begin{equation}
v(t) = \frac{\zeta (t)}{\sqrt{\hbar}} e^{- i \theta (t)},
\end{equation}
and by introducing the momentum $p_{\zeta} = m(t) \dot{\zeta}$,
the effective Hamiltonian (\ref{an c-t ham}) is rewritten as
\begin{eqnarray}
H_{\rm C.T} (t) = \frac{p^2_c}{2 m(t)} + m(t) \frac{\lambda_{2n}
(t) }{(2n)!}q^{2n}_c + \hbar \coth(\frac{\beta \hbar \omega_0}{2})
\Biggl[ \frac{p_{\zeta}^2}{2m (t)} + \frac{1}{8 m(t) \zeta^2}
\Biggr] \nonumber\\ + m(t) \sum_{k = 1}^{n}
\frac{\lambda_{2n}(t)}{2^k k! (2n- 2k)!} \Bigl[ \hbar \zeta^2
\coth(\frac{\beta \hbar \omega_0}{2}) \Bigr]^k q^{2(n-k)}_c.
\end{eqnarray}
Now the equation of motion for the classical field $q_c$ is given
by
\begin{equation}
\ddot{q}_c + \frac{\dot{m}}{m}\dot{q}_c + \frac{\lambda_{2n} (t)
}{(2n-1)!}q^{2n -1}_c + \sum_{k = 1}^{n}
\frac{\lambda_{2n}(t)}{2^k k! (2n- 2k -1)!} [\hbar \zeta^2
\coth(\frac{\beta \hbar \omega_0}{2}) ]^k q^{2n - 2k -1}_c = 0,
\end{equation}
and that for $\zeta$ by
\begin{equation}
\ddot{\zeta} + \frac{\dot{m}}{m} \dot{\zeta} - \frac{1}{4m^2
\zeta^3} + \sum_{k = 1}^{n} \frac{\lambda_{2n}(t)}{2^{k-1} (k-1)!
(2n- 2k)!}  q^{2n - 2k}_c [\hbar \zeta^2 \coth(\frac{\beta
\omega_0}{2}) ]^{k-1} \zeta = 0.
\end{equation}
The phase of $v(t)$ is obtained by the integration
\begin{equation}
\theta (t) = \int \frac{1}{2 m (t)\zeta^2 (t)}.
\end{equation}

\subsection{Coherent State vs. Hartree-Fock Method}

In this subsection we show that the nonequilibrium dynamics
obtained from the effective Hamiltonian (\ref{an c-t ham}) in the
LvN approach recovers exactly the equations of motion from the
mean field and Hartree-Fock methods.

First, the effective Hamiltonian from the coherent state can be
obtained in another way. By dividing $q$ and $p$ into a classical
background and a quantum fluctuation
\begin{equation}
q = q_c + q_f,
\end{equation}
we obtain the expectation value for the Hamiltonian (\ref{an osc})
with respect to the thermal state (\ref{an den op})
\begin{equation}
H_{\rm c.f.}(t) = \frac{p^2_c}{2 m(t)} + \frac{\langle \hat{p}^2_f
\rangle_{\rm T}}{2 m(t)} + m(t) \sum_{k = 0}^{n}
\frac{\lambda_{2n} (t) }{2^k k! (2n- 2k)!} \langle \hat{q}^2_f
\rangle_{\rm T}^k q^{2(n-k)}_c. \label{cl fluct}
\end{equation}
where we have used
\begin{equation}
\langle \hat{q}^{2n} \rangle_{\rm T} = \sum_{k = 0}^{n}
\frac{(2n)!}{2^k k! (2n -2k)!} q_c^{2(n-k)} \Bigl[ \hbar^2 v^* v
\coth(\frac{\beta \hbar \omega_0}{2}) \Bigr]^k,
\end{equation}
and
\begin{equation}
\langle \hat{q}_f^{2n+1} \rangle_{\rm T} = 0 = \langle
\hat{p}_f^{2n+1} \rangle_{\rm T}.
\end{equation}
Noting that $k = 0$ term recovers the classical potential and
\begin{equation}
\langle \hat{p}^2_f \rangle_{\rm T} = m^2 \hbar^2 \dot{v}^*
\dot{v} \coth(\frac{\beta \hbar \omega_0}{2}),
\end{equation}
we can show that Eq. (\ref{cl fluct}) coincides with Eq. (\ref{an
c-t ham}). Therefore the coherent state leads exactly to the
result from the mean-field method.

Second, the Hartree-Fock factorization theorem \cite{cormier1}
leads to the effective Hamiltonian
\begin{equation}
H_{\rm H.F} (t) = \frac{\langle \hat{p}^2 \rangle_{\rm H.F}}{2
m(t)} + m(t) \frac{\lambda_{2n} (t)}{(2n)!} \langle \hat{q}^{2n}
\rangle_{\rm H.F},
\end{equation}
where
\begin{eqnarray}
\langle \hat{p}^{2}\rangle_{\rm H.F} = p_c^2 + \hat{p}^{2}_f,
\nonumber\\ \langle \hat{q}^{2}\rangle_{\rm H.F} = q_c^2 +
\hat{q}^{2}_f
\end{eqnarray}
and
\begin{eqnarray}
\langle \hat{q}^{2n}\rangle_{\rm H.F} &=& \sum_{k = 0}^{2n}
\frac{(2n)!} {k! (2n-k)!} q_c^{2n-k} \hat{q}_f^k \nonumber\\ &=&
\sum_{k = 0}^{n} \frac{(2n)!}{2^k k! (2n-2k)!} q_c^{2(n-k)}
\Bigl[k \langle \hat{q}^{2}_f \rangle_{\rm T}^{k-1} \hat{q}_f^2 -
(k-1) \langle \hat{q}^{2}_f \rangle_{\rm T}^k \Bigr] \nonumber\\
&& + \sum_{k = 0}^{n-1} \frac{(2n)!}{2^k k! (2n-2k-1)!}
q_c^{2n-2k-1} \langle \hat{q}^{2}_f \rangle_{\rm T}^{k} \hat{q}_f.
\end{eqnarray}
for $n \geq 2$. The thermal expectation value of the equation of
motion for $q_c$ yields
\begin{equation}
\ddot{q}_c + \frac{\dot{m}}{m} \dot{q}_c + \sum_{k = 0}^{n}
\frac{\lambda_{2n} (t)}{2^k k! (2n-2k)!} \langle \hat{q}^{2}_f
\rangle_{\rm T}^k q_c^{2n - 2k-1} = 0, \label{hf eq}
\end{equation}
and for $\hat{q}_f$
\begin{equation}
\ddot{\hat{q}}_f + \frac{\dot{m}}{m} \dot{\hat{q}}_f + \sum_{k =
0}^{n} \frac{\lambda_{2n} (t)}{2^{k-1} (k-1)! (2n-2k)!} \langle
\hat{q}^{2}_f \rangle_{\rm T}^{k-1} q_c^{2(n -k)} \hat{q}_f = 0.
\label{hf eq2}
\end{equation}
Therefore it has been shown that Eqs. (\ref{hf eq}) and (\ref{hf
eq2}) are the same as Eqs. (\ref{an osc eq}) and (\ref{an osc
eq2}) from the coherent state representation.

\section{Free Scalar Field for Phase Transition}

As a simple field model for the second order phase transition, we
consider a free complex scalar field, the mass of which changes
the sign during the quench.\footnote{The complex scalar field
model may be related with the time-dependent Landau-Ginsburg
theory provided that the free energy be interpreted as the
Hamiltonian in this paper \cite{rivers}.} The system is described
by the Lagrangian density
\begin{equation}
{\cal L} ({\bf x}, t) = \dot{\Phi}^* ({\bf x}, t) \dot{\Phi} ({\bf
x}, t) - \nabla \Phi^* ({\bf x}, t) \cdot \nabla \Phi ({\bf x}, t)
- m^2(t) \Phi^* ({\bf x}, t) \Phi ({\bf x}, t). \label{lag den1}
\end{equation}
Here the coupling parameter $m^2 (t)$ is assumed to begin with an
initial positive value before, to change the sign during, and to
reach a final negative value after the quench. The $\Phi$ and
$\Phi^*$ are treated as independent fields. The Hamiltonian is
given by
\begin{equation}
H(t) = \int d^3{\bf x} \Bigl[\Pi^* ({\bf x}, t) \Pi ({\bf x}, t) +
\nabla \Phi^* ({\bf x}, t) \cdot \nabla \Phi ({\bf x}, t) + m^2
(t) \Phi^* ({\bf x}, t) \Phi ({\bf x}, t) \Bigr],
\end{equation}
where
\begin{eqnarray}
\Pi ({\bf x}, t) = \frac{\delta {\cal L}({\bf x}, t)}{\delta
\dot{\Phi} ({\bf x}, t)} = \dot{\Phi}^* ({\bf x}, t), \nonumber\\
\Pi^* ({\bf x}, t) = \frac{\delta {\cal L}({\bf x}, t)}{\delta
\dot{\Phi}^*({\bf x}, t)} = \dot{\Phi} ({\bf x}, t)
\end{eqnarray}
are conjugate momenta.

The field and momentum are Fourier-decomposed as
\begin{eqnarray}
\Phi ({\bf x}, t) &=& \int \frac{d^3{\bf k}}{(2\pi)^3} \phi_{\bf
k} (t) e^{i {\bf k} \cdot {\bf x}}, \nonumber\\ \Pi ({\bf x}, t)
&=& \dot{\Phi}^* ({\bf x}, t) = \int \frac{d^3 {\bf k}}{(2\pi)^3}
\dot{\phi}_{\bf k}^* (t) e^{- i {\bf k} \cdot {\bf x}} \equiv \int
\frac{d^3 {\bf k}}{(2\pi)^3} \pi_{\bf k} (t) e^{- i {\bf k} \cdot
{\bf x}}, \label{fourier1}
\end{eqnarray}
and their conjugates as
\begin{eqnarray}
\Phi^* ({\bf x}, t) &=& \int \frac{d^3{\bf k}}{(2\pi)^3} \phi_{\bf
k}^* (t) e^{- i {\bf k} \cdot {\bf x}}, \nonumber\\ \Pi^* ({\bf
x}, t) &=& \dot{\Phi} ({\bf x}, t) = \int \frac{d^3 {\bf
k}}{(2\pi)^3} \dot{\phi}_{\bf k} (t) e^{i {\bf k} \cdot {\bf x}}
\equiv \int \frac{d^3 {\bf k}}{(2\pi)^3} \pi^*_{\bf k} (t) e^{ i
{\bf k} \cdot {\bf x}}.\label{fourier2}
\end{eqnarray}
So space integrals of quadratic fields and momenta result in
momentum integrals for the decoupled modes
\begin{eqnarray}
\int d^3 {\bf x} \Phi^* \Phi &=& \int \frac{d^3 {\bf k}
}{(2\pi)^3} \phi_{\bf k}^* \phi_{\bf k}, \nonumber\\\int d^3 {\bf
x} \Pi^* \Pi &=& \int \frac{d^3 {\bf k} }{(2\pi)^3} \pi_{\bf k}^*
\pi_{\bf k}, \nonumber\\ \int d^3 {\bf x} \nabla \Phi^* \cdot
\nabla \Phi &=& \int \frac{d^3 {\bf k} }{(2\pi)^3} {\bf k}^2
\phi_{\bf k}^* \phi_{\bf k}.
\end{eqnarray}
One then obtains the Hamiltonian as the sum of infinite number of
time-dependent harmonic oscillators
\begin{equation}
H (t) = \int \frac{d^3{\bf k}}{(2\pi)^3} \Bigl[\pi_{\bf k}^*
\pi_{\bf k} + \Bigl({\bf k}^2 + m^2(t)\Bigr)\phi_{\bf k}^*
\phi_{\bf k} \Bigr].
\end{equation}
Canonical quantization is prescribed by imposing the commutation
relations at equal times
\begin{eqnarray}
\bigl[\hat{\phi}_{{\bf k}'} (t), \hat{\pi}_{\bf k} (t) \bigr] = i
\hbar \delta_{{\bf k} , {\bf k}'}, \nonumber\\ \bigl[
\hat{\phi}^*_{{\bf k}'} (t), \hat{\pi}^*_{\bf k} (t) \bigr] = i
\hbar \delta_{{\bf k}, {\bf k}'},
\end{eqnarray}
and all the other commutators vanish. Following Sec. III, we find
the two pairs of the annihilation and creation operators
(\ref{an-cr}) for each ${\bf k}$-mode,
\begin{eqnarray}
\hat{a}_{\bf k}  (t) &=& i \Bigl( \varphi^*_{\bf k} (t)
\hat{\pi}^*_{\bf k} - \dot{\varphi}^*_{\bf k} (t) \hat{\phi}_{\bf
k} \Bigr), \nonumber\\ \hat{a}_{\bf k}^{\dagger} (t) &=& - i
\Bigl( \varphi_{\bf k} (t) \hat{\pi}_{\bf k} - \dot{\varphi}_{\bf
k} (t) \hat{\phi}_{\bf k}^* \Bigr), \label{6-pair1}
\end{eqnarray}
and
\begin{eqnarray}
\hat{a}_{\bf k}^*  (t) &=& i \Bigl( \varphi^*_{\bf k} (t)
\hat{\pi}_{\bf k} - \dot{\varphi}^*_{\bf k} (t) \hat{\phi}^*_{\bf
k} \Bigr), \nonumber\\ \hat{a}_{\bf k}^{*\dagger}  (t) &=& - i
\Bigl( \varphi_{\bf k} (t) \hat{\pi}^*_{\bf k} -
\dot{\varphi}_{\bf k} (t) \hat{\phi}_{\bf k} \Bigr),
\label{6-pair2}
\end{eqnarray}
where $\varphi_{\bf k}$ and $\varphi_{\bf k}^*$ satisfy the same
classical equation of motion
\begin{equation}
\ddot{\varphi}_{\bf k} (t) + \Bigl( {\bf k}^2 + m^2 (t) \Bigr)
\varphi_{\bf k} (t) = 0. \label{fr cl eq}
\end{equation}
They further satisfy the standard commutation relations
\begin{equation}
[\hat{a}_{\bf k'}, \hat{a}^{\dagger}_{\bf k}] = \delta_{{\bf k},
{\bf k}'}, \quad [\hat{a}^*_{\bf k'}, \hat{a}^{*\dagger}_{\bf k}]
= \delta_{{\bf k}, {\bf k}'}.
\end{equation}

The Fock space for each mode can be constructed according to Sec.
III. We consider two symmetric states: the vacuum and thermal
states. The vacuum is the one annihilated by all the $\hat{a}_{\bf
k}$ and $\hat{a}^*_{\bf k}$:
\begin{equation}
\hat{a}_{\bf k} (t) \vert 0, t \rangle = 0, \quad \hat{a}^*_{\bf
k} (t) \vert 0, t \rangle = 0.
\end{equation}
By inverting Eqs. (\ref{6-pair1}) and (\ref{6-pair2}) one
expresses the fields as
\begin{eqnarray}
\hat{\phi}_{\bf k} = \hbar \Bigl(\varphi_{\bf k} \hat{a}_{\bf k} +
\dot{\varphi_{\bf k}}^* \hat{a}_{\bf k}^{*\dagger} \Bigr),
\nonumber\\ \hat{\phi}_{\bf k}^* = \hbar \Bigl(\varphi_{\bf k}
\hat{a}_{*\bf k} + \dot{\varphi_{\bf k}}^* \hat{a}_{\bf
k}^{\dagger} \Bigr),
\end{eqnarray}
from which follow the vacuum expectation values
\begin{eqnarray}
\langle \hat{\Phi}^* \hat{\Phi} \rangle_{\rm V} = \int \frac{d^3
{\bf k}}{(2\pi)^3} \Bigl[ \hbar^2 \varphi^*_{\bf k} (t)
\varphi_{\bf k} (t) \Bigr], \nonumber\\ \langle \hat{\Pi}^*
\hat{\Pi} \rangle_{\rm V} = \int \frac{d^3 {\bf k}}{(2\pi)^3}
\Bigl[ \hbar^2 \dot{\varphi}^*_{\bf k} (t) \dot{\varphi}_{\bf k}
(t) \Bigr].
\end{eqnarray}
The initial thermal state defined by the density operator for each
mode
\begin{eqnarray}
\hat{\rho} (t) = \prod_{\bf k} \hat{\rho}_{\bf k} (t) = \prod_{\bf
k} \Biggl\{\frac{1}{Z_{\bf k}} \exp \Biggl[ - \beta \hbar
\omega_{i, {\bf k}} \Bigl(\hat{a}_{\bf k}^{\dagger} (t)
\hat{a}_{\bf k} (t) + \frac{1}{2} \Bigr) \Biggr] \nonumber\\
\times \frac{1}{Z_{\bf k}^*} \exp \Biggl[ - \beta \hbar \omega_{i,
{\bf k}} \Bigl(\hat{a}_{\bf k}^{*\dagger} (t) \hat{a}_{\bf k}^*
(t) + \frac{1}{2} \Bigr) \Biggr] \Biggr\}
\end{eqnarray}
leads to the thermal expectation values
\begin{eqnarray}
\langle \hat{\Phi}^* \hat{\Phi} \rangle_{\rm T} &=&  {\rm Tr}
\Bigl[\hat{\rho} (t) \hat{\Phi}^* \hat{\Phi} \Bigr] = \int
\frac{d^3 {\bf k}}{(2\pi)^3} \Bigl[ \hbar^2 \varphi^*_{\bf k} (t)
\varphi_{\bf k} (t) \coth (\frac{\beta \hbar \omega_{i, {\bf
k}}}{2}) \Bigr], \nonumber\\ \langle \hat{\Pi}^* \hat{\Pi}
\rangle_{\rm T} &=& {\rm Tr} \Bigl[\hat{\rho} (t) \hat{\Pi}^*
\hat{\Pi} \Bigr] = \int \frac{d^3 {\bf k}}{(2\pi)^3} \Bigl[
\hbar^2 \dot{\varphi}^*_{\bf k} (t) \dot{\varphi}_{\bf k} (t)
\coth (\frac{\beta \hbar \omega_{i, {\bf k}}}{2}) \Bigr].
\end{eqnarray}
One then finds the two-point correlation functions at equal times
by taking the expectation value with respect to the vacuum state
\begin{equation}
G_{\rm V} ({\bf y}, {\bf x}, t) = \langle \hat{\Phi}^* ({\bf y},
t) \hat{\Phi} ({\bf x}, t) \rangle_{\rm V} = \int
\frac{d^3k}{(2\pi)^3} \Bigl[\hbar^2 \varphi^*_{{\bf k}} (t)
\varphi_{{\bf k}} (t) \Bigr] e^{i {\bf k} \cdot ({\bf x} - {\bf
y})}, \label{vac corr}
\end{equation}
and with respect to the thermal state
\begin{equation}
G_{\rm T} ({\bf y}, {\bf x}, t) = \langle \hat{\Phi}^* ({\bf y},
t) \hat{\Phi} ({\bf x}, t) \rangle_{\rm T} = \int
\frac{d^3k}{(2\pi)^3} \Bigl[ \hbar^2 \varphi^*_{\bf k} (t)
\varphi_{\bf k} (t) \coth (\frac{\beta \hbar \omega_{i, {\bf
k}}}{2}) \Bigr] e^{i {\bf k} \cdot ({\bf x} - {\bf y})},
\label{ther corr}
\end{equation}
where
\begin{equation}
\omega_{i, {\bf k}} = \sqrt{{\bf k}^2 + m^2 (- \infty)}.
\end{equation}

\subsection{Instantaneous Quench}

The instantaneous quench model is an analytically solvable one, in
which the mass changes as
\begin{eqnarray}
m^2 (t) = \cases{ m_i^2, & $ t < 0$,  \cr  - m_f^2, & $t > 0$.
\cr} \label{inst freq}
\end{eqnarray}
Before the quench $(t < 0)$, the solution to Eq. (\ref{fr cl eq}),
which also satisfies the condition ({\ref{boun con}), is given by
\begin{equation}
\varphi_{i, {\bf k}} (t) = \frac{1}{\sqrt{2 \hbar \omega_{i, {\bf
k}}}} e^{-i \omega_{i, {\bf k}} t}, \quad \omega_{i, {\bf k}} =
\sqrt{k^2 + m_i^2}. \label{fr sol1}
\end{equation}
Then the two-point vacuum correlation function (\ref{vac corr})
becomes
\begin{eqnarray}
G_{i, {\rm V}} ({\bf y}, {\bf x}, t) &=& \int
\frac{d^3k}{(2\pi)^3} \frac{\hbar}{2 \sqrt{k^2 + m_i^2}}e^{i {\bf
k} \cdot ({\bf x} - {\bf y})} \nonumber\\ &=& \frac{\hbar}{4 \pi^2
} \frac{m_i K_1(m_i |{\bf x} - {\bf y}|)}{|{\bf x} - {\bf y}|},
\label{fr cor1}
\end{eqnarray}
where $K_1$ is the modified Bessel function. Equation (\ref{fr
cor1}) coincides with the result for a massive scalar field in
Ref. \cite{bogoliubov}. Similarly, the two-point thermal
correlation function is given by
\begin{eqnarray}
G_{i, {\rm T}} ({\bf y}, {\bf x}, t) &=& G_{i, {\rm V}} ({\bf y},
{\bf x}, t) \nonumber\\ &+& \frac{\hbar}{2 \pi^2 }
\frac{m_i}{\sqrt{|{\bf x} - {\bf y}|^2 + m_i^2}} \sum_{n =
1}^{\infty} K_1(m_i \sqrt{|{\bf x} - {\bf y}|^2 + (\beta \hbar
n)^2)}). \label{fr cor2}
\end{eqnarray}

On the other hand, after the quench $(t > 0)$, the classical
equations of motion are classified into two types: the one from
the long wavelength modes with $k^2 < m_f^2$ has the negative
frequency squared and exhibits an exponential behavior, and the
other from the short wavelength modes with $k^2 > m_f^2$ still has
the positive frequency squared and shows an oscillatory behavior.
Each mode moves under a constant frequency squared before the
quench time but suddenly experiences a potential step in the case
of short wavelengths and a potential barrier in the case of long
wavelengths. There is the analogy between Eq. (\ref{fr cl eq}) and
the scattering problem of quantum mechanics. The solution to Eq.
(\ref{fr cl eq}) together with the initial asymptotic data
(\ref{fr sol1}) is the complex conjugate of the scattering wave
function by either the potential step or barrier \cite{kim7}. The
solution to Eq. (\ref{fr cl eq}) after the quench should match at
the quench time $t = 0$ continuously with Eq. (\ref{fr sol1})
before the quench. It is rather straightforward to find such
solutions for the short wavelength modes $(k^2 > m_f^2)$
\begin{equation}
\varphi_{f_S, {\bf k}} (t) = \frac{1}{\sqrt{2 \hbar \omega_{i,
{\bf k}}}} \Biggl[- i \frac{\omega_{i, {\bf k}}}{\omega_{f, {\bf
k}}} \sin( \omega_{f, {\bf k}} t) + \cos ( \omega_{f, {\bf k}} t)
\Biggr], \quad \omega_{f, {\bf k}} = \sqrt{k^2 - m_f^2}, \label{fr
sol2}
\end{equation}
and for the long wavelength modes $(k^2 < m_f^2)$
\begin{equation}
\varphi_{f_U, {\bf k}} (t) = \frac{1}{\sqrt{2 \hbar \omega_{i,
{\bf k}}}} \Biggl[- i \frac{\omega_{i, {\bf
k}}}{\tilde{\omega}_{f, {\bf k}}} \sinh( \tilde{\omega}_{f, {\bf
k}} t) + \cosh ( \tilde{\omega}_{f, {\bf k}} t) \Biggr], \quad
\tilde{\omega}_{f, {\bf k}} = \sqrt{m_f^2 - k^2}. \label{fr sol3}
\end{equation}

A few comments are in order. The solution (\ref{fr sol3})
represents an instability due to the phase transition and is
obtained by continuing analytically the solution (\ref{fr sol2}).
When Eq. (\ref{fr sol2}) is rewritten as
\begin{equation}
\varphi_{f, {\bf k}} (t) = \frac{1}{\sqrt{2 \hbar \omega_{i, {\bf
k}}}} \Biggl[\Biggl(\frac{ \omega_{f, {\bf k}} + \omega_{i, {\bf
k}}}{2 \omega_{f, {\bf k}}} \Biggr) e^{- i \omega_{f, {\bf k}} t}
+ \Biggl( \frac{\omega_{f, {\bf k}} - \omega_{i, {\bf
k}}}{2\omega_{f, {\bf k}}} \Biggr) e^{ i \omega_{f, {\bf k}} t}
\Biggr], \label{fr sol4}
\end{equation}
the first and second terms correspond to the positive and negative
frequencies, respectively, hence the second term explains the
particle creation by changing frequency \cite{birrel}.

After some manipulations of algebra, we obtain the two-point
vacuum correlation function after the quench
\begin{eqnarray}
G_{f, {\rm V}} ({\bf y}, {\bf x}, t) &=& G_{i, {\rm V}} ({\bf y},
{\bf x}, t)\nonumber\\ &+& \frac{\hbar}{4 \pi^2 |{\bf x} - {\bf
y}|}\int_{0}^{m_f} dk k \Biggl(\frac{\omega^2_{i, {\bf k}} +
\tilde{\omega}^2_{f, {\bf k}}}{\omega_{i, {\bf
k}}\tilde{\omega}^2_{f, {\bf k}}} \Biggr) \sin(k |{\bf x} - {\bf
y}| ) \sinh^2 (\tilde{\omega}_{f, {\bf k}} t) \nonumber\\ &+&
 \frac{\hbar}{4 \pi^2 |{\bf x} - {\bf
y}|}\int_{m_f}^{\infty} dk k \Biggl(\frac{\omega^2_{i, {\bf k}} -
\omega^2_{f, {\bf k}}}{\omega_{i, {\bf k}}\omega^2_{f, {\bf k}}}
\Biggr) \sin(k |{\bf x} - {\bf y}| ) \sin^2 (\omega_{f, {\bf k}}
t). \label{vac cor}
\end{eqnarray}
Similarly, the two-point thermal correlation function is given by
\begin{eqnarray}
G_{f, {\rm T}} ({\bf y}, {\bf x}, t) &=& G_{i, {\rm T}}({\bf y},
{\bf x}, t)\nonumber\\ &+& \frac{\hbar}{4 \pi^2 |{\bf x} - {\bf
y}|}\int_{0}^{m_f} dk k \Biggl(\frac{\omega^2_{i, {\bf k}} +
\tilde{\omega}^2_{f, {\bf k}}}{\omega_{i, {\bf
k}}\tilde{\omega}^2_{f, {\bf k}}} \Biggr) \sin(k |{\bf x} - {\bf
y}| ) \sinh^2 (\tilde{\omega}_{f, {\bf k}} t) \coth
\Bigl(\frac{\beta \hbar \omega_{i, {\bf k}}}{2} \Bigr) \nonumber\\
&+&
 \frac{\hbar}{4 \pi^2 |{\bf x} - {\bf
y}|}\int_{m_f}^{\infty} dk k \Biggl(\frac{\omega^2_{i, {\bf k}} -
\omega^2_{f, {\bf k}}}{\omega_{i, {\bf k}}\omega^2_{f, {\bf k}}}
\Biggr) \sin(k |{\bf x} - {\bf y}| ) \sin^2 (\omega_{f, {\bf k}}
t) \coth \Bigl(\frac{\beta \hbar \omega_{i, {\bf k}}}{2} \Bigr).
\label{ther cor}
\end{eqnarray}
The first terms in Eqs. (\ref{vac cor}) and (\ref{ther cor}) are
the two-point vacuum and thermal correlations (\ref{fr cor1}) and
(\ref{fr cor2}), respectively, before the quench. Therefore the
remaining two terms describe the effect of the quench. In
particular, the second terms are dominant and rooted on the
instability during the phase transition, which is missing in the
field theoretical approach to equilibrium dynamics. Note that
$\omega^2_{i, {\bf k}} - \omega^2_{f, {\bf k}} = m_i^2 + m_f^2$
and $\omega^2_{i, {\bf k}} + \tilde{\omega}^2_{f, {\bf k}} = m_i^2
+ m_f^2$, so the amplitudes of $ \sin(k |{\bf x} - {\bf y}|)$
decrease as $1/k^3$ for very short wavelengths, hence short
wavelengths contribute negligibly. However, there is a residual
contribution from near the critical wavelength $k_c = m_f$, which
becomes much smaller than the second terms at later times and will
not be considered any more.

We wish to determine the size of domains from the second order
phase transition of the instantaneous quench. At later times $(m_f
t \gg 1)$ after the quench the dominant contribution to Eq.
(\ref{ther cor}) comes from the second term, so one has
approximately
\begin{equation}
G_{f_U, {\rm T}} (r, t) \simeq  \frac{\hbar}{16 \pi^2 r
}\int_{0}^{m_f} dk \Bigl\{k e^{2 \tilde{\omega}_{f, {\bf k}} t}
\Bigr\} \sin(rk) F_{\rm I} (k),  \label{st int}
\end{equation}
where $r = |{\bf x} - {\bf y}|$, and
\begin{equation}
F_{\rm I}  (k) = \Biggl(\frac{m_i^2 + m_f^2}{\omega_{i, {\bf
k}}\tilde{\omega}^2_{f, {\bf k}}} \Biggr) \coth \Bigl(\frac{\beta
\hbar \omega_{i, {\bf k}}}{2} \Bigr).
\end{equation}
The function of $k$ in the curly bracket of Eq. (\ref{st int}) has
a sharp peak at $k_0 = \sqrt{m_f/2t}$, whereas $F_{\rm I} (k)$ is
a slowly varying function. We employ the steepest descent method
(see Appendix C) to obtain
\begin{equation}
G_{f_U, {\rm T}} (r, t) \simeq G_{f_U, {\rm T}} (0, t)
\frac{\sin\Bigl(\sqrt{\frac{m_f}{2t}}r\Bigr)}{\sqrt{\frac{m_f}{2t}r}}
\exp \Bigl[-\frac{m_f r^2}{8 t } \Bigr],
\end{equation}
where
\begin{equation}
G_{f_U, {\rm T}} (0, t) = \frac{\hbar}{16
\pi^2}\sqrt{\frac{\pi}{2e}} \Bigl(\frac{m_f}{2t} \Bigr)^{3/2}
e^{2m_ft} F_{\rm I} \Bigl(k_0 = \sqrt{\frac{m_f}{2t}}\Bigr).
\end{equation}
Therefore the size of domains grows according to the classical
Cahn-Allen scaling relation \cite{bray}
\begin{equation}
\xi_D (t) = \sqrt{\frac{8 t}{m_f}}. \label{cahn}
\end{equation}
The scaling relation (\ref{cahn}) for the instantaneous quench
confirms the result obtained in Refs. \cite{boyanovsky,bowick}.

\subsection{Finite  Smooth Quench}

The instantaneous quench does not exhibit the essential spinodal
behavior during the quench process. To see the dynamics of the
second order phase transition one needs a finite quench period.
Such a finite and smooth quench model is described by a field with
the mass given by
\begin{equation}
m^2 (t) = - m_1^2 - m_0^2 \tanh \Bigl(\frac{t}{\tau} \Bigr), \quad
(m_0^2 > |m_1^2|). \label{sm freq}
\end{equation}
At earlier times $t =  - \infty$, the mass has the initial value
\begin{equation}
m^2 = m_i^2 = m_0^2 - m_1^2 > 0,
\end{equation}
and at later times $t = \infty$, the final value
\begin{equation}
m^2 = - m_f^2 = - (m_0^2 + m_1^2) < 0.
\end{equation}
Here $\tau$ measures the quench rate: the large $\tau$-limit
implies that the mass changes slowly from $m^2_i$ at $t = -
\infty$ to $- m_f^2$ at $t = + \infty$, whereas the small
$\tau$-limit implies a rapid change of the mass. In particular,
the $(\tau = 0)$-limit corresponds to the instantaneous change
from $m_i^2$ to $- m_f^2$ at $t = 0$. That is, the instantaneous
quench is a special case of the finite smooth quench model. To
find the Fock space for each mode one needs to solve the classical
equation of motion
\begin{equation}
\ddot{\varphi}_{\bf k} (t) + \Bigl({\bf k}^2 - m_1^2 - m_0^2 \tanh
\Bigl(\frac{t}{\tau} \Bigr) \Bigr) \varphi_{\bf k} (t) = 0.
\label{sm cl eq}
\end{equation}
It should be noted that, as in the instantaneous quench model,
long wavelength modes $(k \leq k_c = m_f)$ let the frequency
change the sign at later times $(t \gg \tau)$
\begin{equation}
\omega_{\bf k}^2 (t) = {\bf k}^2 - m_1^2 - m_0^2 < 0,
\end{equation}
and suffer from the spinodal instability. Each long wavelength
mode has a different quench time determined by $ \omega_{\bf k}
(t_{\bf k}) = 0$.

The solutions to Eq. (\ref{sm cl eq}) are found separately for the
stable modes and unstable modes. The stable modes $(k \geq m_f)$
have the solutions
\begin{equation}
\varphi_{\bf k} (t) = C_{\bf k} e^{2 p_{\bf k} t} {}_2F_1
(\beta_{+, {\bf k}}, \beta_{-, {\bf k}}; \gamma_{\bf k}; -
e^{2t/\tau}), \label{sm sol1}
\end{equation}
where
\begin{eqnarray}
p_{\bf k} &=& - i \frac{1}{2} \omega_{i, {\bf k}}  , \nonumber\\
\beta_{{\pm},{\bf k}} &=& - i \frac{\tau}{2} \Bigl(\omega_{i, {\bf
k}} \pm \omega_{f, {\bf k}} \Bigr), \nonumber\\ \gamma_{\bf k} &=&
1 - i \tau \omega_{i, {\bf k}},
\end{eqnarray}
with
\begin{equation}
\omega_{i, {\bf k}} = \sqrt{k^2+ m_i^2}, \quad \omega_{f, {\bf k}}
= \sqrt{k^2 - m_f^2}.
\end{equation}
Whereas the unstable modes $(k < m_f)$ have the solutions
\begin{equation}
\varphi_{\bf k} (t) = C_{\bf k} e^{2 p_{\bf k} t} {}_2F_1
(\tilde{\beta}_{+, {\bf k}}, \tilde{\beta}_{-, {\bf k}};
\gamma_{\bf k}; - e^{2t/\tau}), \label{sm sol2}
\end{equation}
where
\begin{equation}
\tilde{\beta}_{{\pm},{\bf k}} = - \frac{\tau}{2} \Bigl(i
\omega_{i, {\bf k}} \pm \tilde{\omega}_{f, {\bf k}} \Bigr),
\end{equation}
with
\begin{equation}
\tilde{\omega}_{f, {\bf k}} = \sqrt{m_f^2 - k^2}.
\end{equation}
At earlier times $(\tau \ll - \tau)$ before the quench begins,
both the solutions (\ref{sm sol1}) and (\ref{sm sol2}) have the
same asymptotic form
\begin{equation}
\varphi_{i, {\bf k}} (t) = C_{\bf k} e^{-i \omega_{i, {\bf k}} t},
\label{sm in sol}
\end{equation}
so the constant is fixed to satisfy Eq. (\ref{boun con})
\begin{equation}
c_{\bf k} = \frac{1}{\sqrt{2 \hbar \omega_{i, {\bf k}}}}.
\end{equation}

\subsubsection{During the Quench}

During the quench process $(|t| < \tau)$, the asymptotic forms for
the solutions (\ref{sm sol1}) and (\ref{sm sol2}) are
unfortunately not available. Instead, one may expand the mass
(\ref{sm freq}) to the linear order
\begin{equation}
m^2 (t) = - m_1^2 - m_0^2 \Bigl(\frac{t}{\tau}\Bigr), \label{sm
lin freq}
\end{equation}
which is a good approximation as far as $|t| \ll \tau$. But we
assume $m_0 \tau \gg 1$ so that the linear quench process is
sufficiently long enough to allow the asymptotic analysis. Each
mode of the scalar field then has the frequency squared
\begin{equation}
\omega^2_{\bf k} (t) = {\bf k}^2 - m_1^2 - m_0^2
\Bigl(\frac{t}{\tau} \Bigr) = m_0^2 \Biggl(\frac{t_{\bf k} -
t}{\tau} \Biggr), \label{lin freq}
\end{equation}
where $t_{\bf k}$ is the quench time for the corresponding
unstable mode
\begin{equation}
t_{\bf k} = \frac{\tau}{m_0^2} ({\bf k}^2 - m_1^2).
\end{equation}
Unless $|t_{\bf k}/\tau| \ll 1$, the quench time $t_{\bf k}$
occurs outside the valid regime for Eq. (\ref{sm lin freq}). So we
restrict our attention to those unstable modes with $t_{\bf k} \ll
\tau$. The frequency (\ref{lin freq}) corresponds to the special
case of Eq. (\ref{gen freq}), in which $l_1 = l_2 = 0$, $t_0 =
t_{\bf k}$, $t_0 - t_{i} = \tau$ and $\omega_i = m_0$. Then the
solution in the intermediate regime before the quench, matching
with the initial solution (\ref{sm in sol}), is given by Eq.
(\ref{gen sol}):
\begin{equation}
\varphi_{m_S, {\bf k}} (t) = D_1 z^{1/3}_{\bf k} H^{(2)}_{1/3}
(z_{\bf k}) + D_2 z^{1/3}_{\bf k} H^{(1)}_{1/3} (z_{\bf k}),
\label{lin sol}
\end{equation}
where
\begin{equation}
z_{\bf k} = \frac{2}{3} m_0 \tau \Biggl(\frac{t_{\bf k} - t}{\tau}
\Biggr)^{1/3}.
\end{equation}
The coefficients $D_1$ and $D_2$ are given by Eq. (\ref{gen
coef}). After the quench time $ (t > t_{\bf k})$ for each mode,
the solution (\ref{lin sol}) is analytically continued for the
unstable mode
\begin{eqnarray}
\varphi_{m_U, {\bf k}} (t) = \frac{D_1}{2} \tilde{z}^{1/3}_{\bf k}
\Biggl[ e^{i \frac{3}{2}\pi} H^{(2)}_{1/3} (i \tilde{z}_{\bf k}) +
e^{i \frac{1}{2}\pi} H^{(2)}_{1/3} (-i \tilde{z}_{\bf k}) \Biggr]
\nonumber\\ +  \frac{D_2}{2} \tilde{z}^{1/3}_{\bf k} \Biggl[ e^{i
\frac{3}{2}\pi} H^{(1)}_{1/3} (i \tilde{z}_{\bf k}) + e^{i
\frac{1}{2}\pi} H^{(1)}_{1/3} (-i \tilde{z}_{\bf k}) \Biggr],
\label{lin un sol}
\end{eqnarray}
where
\begin{equation}
\tilde{z}_{\bf k} = \frac{2}{3} m_0 \tau \Biggl(\frac{t - t_{\bf
k}}{\tau} \Biggr)^{1/3}.
\end{equation}
The two-point thermal correlation function
\begin{equation}
G_{m, {\rm T}} (r, t) = \frac{\hbar^2}{2 \pi^2}\int_{0}^{m_f} dk
k^2  \frac{\sin(kr)}{kr} \varphi^*_{m_U, {\bf k}} (t)
\varphi_{m_U, {\bf k}} (t) + \frac{\hbar^2}{2
\pi^2}\int_{m_f}^{\infty} dk k^2 \frac{\sin(kr)}{kr}
\varphi^*_{m_S, {\bf k}} (t) \varphi_{m_S, {\bf k}} (t), \label{sm
corr}
\end{equation}
with $r = |{\bf x} - {\bf y}|$, is dominated by the unstable modes
(\ref{lin un sol}), since the stable modes (\ref{lin sol})
oscillate rapidly and do contribute little. By using the
asymptotic form for the solution (\ref{lin un sol}) in the regime
$\tau \gg t \gg (\tau/m_0^2)^{1/3}$
\begin{equation}
\varphi_{m_U, {\bf k}} (t) = \sqrt{\frac{1}{2\pi}}
\frac{e^{\tilde{z}_{\bf k}}}{\tilde{z}_{\bf k}^{1/6}} \Bigl[ e^{-i
\frac{1}{3}\pi}D_1 + e^{- i \frac{2}{3}\pi} D_2 \Bigr]. \label{lin
un asym}
\end{equation}
one has approximately
\begin{equation}
\varphi^*_{m_U, {\bf k}} (t) \varphi_{m_U, {\bf k}} (t) =
\frac{1}{8 \hbar} \Biggl\{\frac{1}{m_0}
\Bigl(\frac{z_i}{\tilde{z}_{\bf k}} \Bigr)^{1/3} \sin^2 \Bigl(z_i
+ \frac{\pi}{4} \Bigr) \Biggr\} e^{2 \tilde{z}_{\bf k}},
\end{equation}
where $ z_i = 2m_0 \tau / 3$. So the correlation function takes
the form
\begin{equation}
G_{m_U, {\rm T}} (r, t) \simeq  \frac{\hbar}{16 \pi^2 r
}\int_{0}^{m_f} dk \Bigl\{k e^{2 \tilde{\omega}_{f, {\bf k}}
\tilde{t}} \Bigr\} \sin(kr) F_{\rm II} (k), \label{sm corr2}
\end{equation}
where
\begin{equation}
F_{\rm II} (k) = \frac{1}{m_0} \Bigl(\frac{\frac{2}{3} m_0
\tau}{\tilde{z}_{\bf k}} \Bigr)^{1/3} \sin^2 \Bigl(\frac{2}{3} m_0
\tau + \frac{\pi}{4} \Bigr) \coth \Bigl(\frac{\beta \hbar
\omega_{i, {\bf k}}}{2} \Bigr).
\end{equation}
The function in the curly bracket is rapidly varying and has a
peak at $k_0 = (m_0/2 \sqrt{\tau t})^{1/2}$. Hence after applying
the steepest decent method (see Appendix C), we finally obtain
\begin{equation}
G_{m_U, {\rm T}} (r, t) \simeq G_{m_U, {\rm T}} (0, t)
\frac{\sin\Bigl(\frac{\sqrt{\tau t}}{m_0}r\Bigr)}{\frac{\sqrt{\tau
t}}{m_0}r} \exp \Bigl[-\frac{r^2}{8 \frac{\sqrt{\tau t}}{m_0} }
\Bigr],
\end{equation}
where
\begin{equation}
G_{m_U, {\rm T}} (0, t) = \frac{\hbar}{64 \pi^2 m_0}
\Biggl(\frac{\pi m_0^3}{(\tau t)^{3/2}} \Biggr)^{1/2}
e^{(\frac{4}{3}m_0 t + 2 \frac{m_1^2}{m_0} \tau )
\sqrt{\frac{t}{\tau}}} F_{II} \Bigl(k_0 = (\frac{m_0}{2 \sqrt{\tau
t}})^{1/2} \Bigr).
\end{equation}
We thus have shown the scaling relation for the domain size
\begin{equation}
\xi_D (t) = 2 \Biggl(\frac{2\tau t}{m_0^2}\Biggr)^{1/4}. \label{m
sc rel}
\end{equation}
The scaling relation (\ref{m sc rel}) confirms, up to a numerical
factor, the result in Ref. \cite{rivers}. However, the power-law
is different from the Cahn-Allen scaling relation after the
completion of quench.

\subsubsection{After the Quench}

At later times $(t \gg \tau)$ after the completion of quench, the
solution (\ref{sm sol1}) has the asymptotic form
\begin{eqnarray}
&& \varphi_{f, {\bf k}} (t) =  \frac{e^{2 p_{\bf k} t}}{\sqrt{2
\hbar \omega_{i, {\bf k}}}}
 \Biggl[ \frac{\Gamma (\gamma_{\bf k}) \Gamma
(\beta_{-, {\bf k}} - \beta_{+, {\bf k}})}{\Gamma(\beta_{-,{\bf
k}} )\Gamma(\gamma_{\bf k} - \beta_{+, {\bf k}})}e^{- 2 \beta_{+,
{\bf k}} t/\tau} {}_2F_1 (\beta_{+, {\bf k}}, 1 - \gamma_{\bf k} +
\beta_{+, {\bf k}}; 1 - \beta_{-, {\bf k}} + \beta_{+, {\bf k}}; -
e^{- 2t/\tau}) \nonumber\\ && ~~~~ + \frac{\Gamma (\gamma_{\bf k})
\Gamma (\beta_{+, {\bf k}} - \beta_{-, {\bf
k}})}{\Gamma(\beta_{+,{\bf k}} )\Gamma(\gamma_{\bf k} - \beta_{-,
{\bf k}})} e^{- 2 \beta_{-, {\bf k}} t/\tau} {}_2F_1 (\beta_{-,
{\bf k}}, 1 - \gamma_{\bf k} + \beta_{-, {\bf k}}; 1 - \beta_{+,
{\bf k}} + \beta_{-, {\bf k}}; - e^{- 2t/\tau}) \Biggr].
\end{eqnarray}
The asymptotic form for Eq. (\ref{sm sol2}) is obtained by
replacing $\beta_{\pm, {\bf k}}$ by $\tilde{\beta}_{\pm, {\bf
k}}$. From the asymptotic form of the hypergeometric function
\cite{abramowitz}, we find the asymptotic form for the stable
modes
\begin{eqnarray}
\varphi_{f_S, {\bf k}} (t) =  \frac{1}{\sqrt{2 \hbar \omega_{i,
{\bf k}}}} \frac{\Gamma (1- i \omega_{i, {\bf k}}\tau) \Gamma (-
i\omega_{f, {\bf k}}\tau)}{\Gamma(1 - i\frac{\tau}{2}(\omega_{i,
{\bf k}}+\omega_{f, {\bf k}}))\Gamma(- i\frac{\tau}{2}(\omega_{i,
{\bf k}}+\omega_{f, {\bf k}}))} e^{-i \omega_{f, {\bf k}} t}
\nonumber\\ + \frac{1}{\sqrt{2 \hbar \omega_{i, {\bf k}}}}
\frac{\Gamma (1-i \omega_{i, {\bf k}}\tau) \Gamma ( i\omega_{f,
{\bf k}}\tau)}{\Gamma(1 - i\frac{\tau}{2}(\omega_{i, {\bf k}} -
\omega_{f, {\bf k}}))\Gamma(- i\frac{\tau}{2}(\omega_{i, {\bf k}}
- \omega_{f, {\bf k}}))} e^{i \omega_{f, {\bf k}} t}, \label{st
asy}
\end{eqnarray}
and for the unstable modes
\begin{eqnarray}
\varphi_{f_U, {\bf k}} (t) =  \frac{1}{\sqrt{2 \hbar \omega_{i,
{\bf k}}}} \frac{\Gamma (1-i \omega_{i, {\bf k}}\tau) \Gamma
(\tilde{\omega}_{f, {\bf k}}\tau)}{(-1) \frac{\tau}{2}(i
\omega_{i, {\bf k}} - \tilde{\omega}_{f, {\bf k}})\Gamma^2(-
\frac{\tau}{2}(i \omega_{i, {\bf k}} - \tilde{\omega}_{f, {\bf
k}}))} e^{\tilde{\omega}_{f, {\bf k}} t} \nonumber\\ +
\frac{1}{\sqrt{2 \hbar \omega_{i, {\bf k}}}} \frac{\Gamma (1 - i
\omega_{i, {\bf k}}\tau) \Gamma (- \tilde{\omega}_{f, {\bf
k}}\tau)}{(-1) \frac{\tau}{2}(i \omega_{i, {\bf k}} +
\tilde{\omega}_{f, {\bf k}})\Gamma^2(- \frac{\tau}{2}(i \omega_{i,
{\bf k}} + \tilde{\omega}_{f, {\bf k}}))} e^{- \tilde{\omega}_{f,
{\bf k}} t}. \label{unst asy}
\end{eqnarray}

Now a few comments are in order. First, the coefficient of the
positive frequency asymptotic solution for each short wavelength
mode
\begin{equation}
\varphi_{\bf k}^{out} (t) = \frac{1}{\sqrt{2 \hbar \omega_{f, {\bf
k}}}}e^{- i \omega_{f, {\bf k}} t}
\end{equation}
leads to the rate for the initial vacuum to remain in the final
vacuum
\begin{equation}
\vert \langle 0_{\bf k}, + \infty \vert 0_{\bf k}, - \infty
\rangle \vert^2 = \frac{\sinh^2 \Bigl[\frac{\pi \tau}{2}\Bigl(
\omega_{i, {\bf k}}+ \omega_{f, {\bf k}}\Bigr) \Bigr]}{\sinh (\pi
\tau \omega_{i, {\bf k}})\sinh (\pi \tau \omega_{f, {\bf k}})}.
\end{equation}
On the other hand, the coefficient of the negative frequency
solution $\varphi_{\bf k}^{out*} (t)$ leads to the particle
production rate \cite{birrel}
\begin{equation}
1 - \vert \langle 0_{\bf k}, + \infty \vert 0_{\bf k}, - \infty
\rangle \vert^2 = \frac{\sinh^2 \Bigl[\frac{\pi \tau}{2}\Bigl(
\omega_{i, {\bf k}}- \omega_{f, {\bf k}}\Bigr) \Bigr]}{\sinh (\pi
\tau \omega_{i, {\bf k}})\sinh (\pi \tau \omega_{f, {\bf k}})}.
\end{equation}
Second, when $\tilde{\omega}_{f {\bf k}} \tau \geq 1$, the
coefficient of the decaying mode in Eq. (\ref{unst asy}) can
become infinite at
\begin{equation}
\tilde{\omega}_{f, {\bf k}} \tau = n, \quad (n = 1, 2, 3 ,
\cdots), \label{res}
\end{equation}
because the gamma function has simple poles at these negative
integers.  For a rapid quench $\tau \ll 1$, there does not exist
any integers that satisfy Eq. (\ref{res}). Hence this kind of
resonance can happen only for a non-negligible $\tau$, {\it i.e.},
for a very slow quench, which will be treated elsewhere. Contrary
to the resonance at (\ref{res}), the apparent singularity at
$\tilde{\omega}_{f, {\bf k}} \tau = 0$ is removed by considering
both terms in Eq. (\ref{unst asy}):
\begin{eqnarray}
\varphi_{f_U {\bf k}} (t) &=& \frac{1}{\sqrt{2 \hbar \omega_{i,
{\bf k}}}} \frac{\Gamma (1-i \omega_{i, {\bf k}}\tau)}{(-i
\frac{\tau}{2} \omega_{i, {\bf k}}) \Gamma^2(-i \frac{\tau}{2}
\omega_{i, {\bf k}})} \Bigl[\Gamma (-\tilde{\omega}_{f, {\bf
k}}\tau) e^{-\tilde{\omega}_{f, {\bf k}} t} + \Gamma
(\tilde{\omega}_{f, {\bf k}}\tau) e^{\tilde{\omega}_{f, {\bf k}}
t} \Bigr] \nonumber\\ &=& \frac{1}{\sqrt{2 \hbar \omega_{i, {\bf
k}}}} \frac{\Gamma (1-i \omega_{i, {\bf k}}\tau)}{(-i
\frac{\tau}{2} \omega_{i, {\bf k}}) \Gamma^2(-i \frac{\tau}{2}
\omega_{i, {\bf k}})} \Bigl[ 2 \frac{\sinh(\tilde{\omega}_{f, {\bf
k}} t)}{\tilde{\omega}_{f, {\bf k}} \tau} \Bigr].
\end{eqnarray}

After the completion of quench $(t \gg \tau)$, the unstable modes
(\ref{unst asy}) dominate the correlation function (\ref{sm corr})
over the stable modes (\ref{st asy}). In particular, the first
term of Eq. (\ref{unst asy}) grows exponentially, so by using the
asymptotic form in Appendix D, one has approximately
\begin{equation}
\varphi^*_{f_U, {\bf k}} (t) \varphi_{f_U, {\bf k}} (t) =
\frac{1}{8 \hbar} \Bigg\{ \Biggl(\frac{\omega_{i, {\bf k}}^2 +
\tilde{\omega}_{f, {\bf k}}^2}{\omega_{i, {\bf k}}
\tilde{\omega}_{f, {\bf k}}} \Biggr)  \frac{\pi \omega_{i, {\bf
k}} \tau}{\sinh(\pi \omega_{i, {\bf k}})} \Biggl[\frac{\Bigl(1 +
\frac{\tau}{2} \tilde{\omega}_{f, {\bf k}} \Bigr)^2 +
\frac{\tau^2}{4}\omega_{i, {\bf k}}}{1 + \tau \tilde{\omega}_{f,
{\bf k}}} \Biggr]^2 e^{\frac{\tau^2}{2} (\zeta(2)-1) (\omega^2_{i,
{\bf k}} + \tilde{\omega}^2_{f, {\bf k}})} \Biggr\} e^{2
\tilde{\omega}_{f, {\bf k}} \tilde{t}},
\end{equation}
where $\zeta (n)$ is the Riemann zeta function and
\begin{equation}
\tilde{t} = t - \frac{\tau^3}{8} \Bigl(\zeta(3) -1\Bigr)
\Bigl(\omega^2_{i, {\bf k}} + \tilde{\omega}^2_{f, {\bf k}}\Bigr).
\label{time lag}
\end{equation}
Note that $\omega^2_{i, {\bf k}} + \tilde{\omega}^2_{f, {\bf k}} =
m_i^2 + m_f^2$, so $\tilde{t}$ lags by a constant in proportion to
the cubic power of the quench period $\tau$. This time-lag is
determined not only by the quench period but also by the initial
and final coupling constants $m_i$ and $m_f$. After applying the
steepest decent method to the correlation function (see Appendix
C), we finally obtain
\begin{equation}
G_{f_U, {\rm T}} (r, t) \simeq G_{f_U, {\rm T}} (0, t)
\frac{\sin\Bigl(\sqrt{\frac{m_f}{2\tilde{t}}}r\Bigr)}{\sqrt{\frac{m_f}{2\tilde{t}}r}}
\exp \Bigl[-\frac{m_f r^2}{8 \tilde{t} } \Bigr],
\end{equation}
where
\begin{equation}
G_{f_U, {\rm T}} (0, t) = \frac{\hbar}{16
\pi^2}\sqrt{\frac{\pi}{2e}} \Bigl(\frac{m_f}{2\tilde{t}}
\Bigr)^{3/2} e^{2m_f\tilde{t}} F_{\rm III} \Bigl(k_0 =
\sqrt{\frac{m_f}{2\tilde{t}}}\Bigr).
\end{equation}
Here
\begin{equation}
F_{\rm III} (k) = \Biggl(\frac{\omega_{i, {\bf k}}^2 +
\tilde{\omega}_{f, {\bf k}}^2}{\omega_{i, {\bf k}}
\tilde{\omega}_{f, {\bf k}}} \Biggr) \frac{\pi \omega_{i, {\bf k}}
\tau}{\sinh(\pi \omega_{i, {\bf k}})} \Biggl[\frac{\Bigl(1 +
\frac{\tau}{2} \tilde{\omega}_{f, {\bf k}} \Bigr)^2 +
\frac{\tau^2}{4}\omega_{i, {\bf k}}}{1 + \tau \tilde{\omega}_{f,
{\bf k}}} \Biggr]^2 e^{\frac{\tau^2}{2} (\zeta(2)-1) (\omega^2_{i,
{\bf k}} + \tilde{\omega}^2_{f, {\bf k}})} \coth \Bigl(\frac{\beta
\hbar \omega_{i, {\bf k}}}{2} \Bigr).
\end{equation}
Remarkably, the scaling relation of the domain size has the same
form as Eq. (\ref{cahn}) from the instantaneous quench
\begin{equation}
\xi_D (t) = \sqrt{\frac{8 \tilde{t}}{m_f}}. \label{cahn2}
\end{equation}
However, there is a definite time-lag due to the finite quench as
claimed in Ref. \cite{bowick}. The scaling relation (\ref{cahn2})
is robust, because only the asymptotic form of the exact solutions
(\ref{sm sol2}) is used.

\section{Back-Reaction in $U(1)$-Theory for Phase Transitions}

The free scalar field model in Sec. VI does not describe a real
system for the second order phase transition because the spinodal
instability continues indefinitely. This model describes more
appropriately an intermediate process of phase transition toward
the spinodal line. In this section we consider $|\Phi|^4$-theory,
in which there is a natural exit for the spinodal decomposition.
The spinodal decomposition ends at the spinodal line, the regime
in which the instability from the quench competes with the
back-reaction from the $|\Phi|^4$-potential term.

The $U(1)$-model for the second order phase transition is
described by the Lagrangian density
\begin{equation}
{\cal L} (t) = \dot{\Phi}^* \dot{\Phi} - \nabla \Phi^*\cdot \nabla
\Phi - \frac{\lambda}{4!} \Biggl(\Phi^* \Phi + \frac{12
m^2(t)}{\lambda} \Biggr)^2, \label{lag den7}
\end{equation}
where $m^2 (t)$ is assumed to take either (\ref{inst freq}) or
(\ref{sm freq}). Hence, the coupling parameter $m^2 (t)$ of the
quadratic term starts with the positive constant value $m_i^2$ at
earlier times before the quench, keeps decreasing across zero at
the moment of quench $t_0$, and finally reaches the negative
constant value $- m_f^2$ at later times after the quench. So the
$\Phi = 0$ remains the true minimum of the potential until $t_0$.
However, $\Phi = 0$ is no longer the true minimum after $t_0$, but
becomes a local maximum. The true minimum occurs at $|\Phi|^2 = 12
m^2(t)/\lambda$, and the system undergoes a second order phase
transition. When the system is in the initial thermal equilibrium
at earlier times, it is invariant under $\Phi \rightarrow \Phi
e^{i \theta}$ and has a global $U(1)$-symmetry. As the system
undergoes the second order phase transition, the symmetry is
broken and there occurs topological defects.

>From the Lagrangian density (\ref{lag den7}) follows the
Hamiltonian density
\begin{equation}
{\cal H} ({\bf x}, t) =  \Pi^* \Pi + \nabla \Phi^* \cdot \nabla
\Phi^* + m^2 (t) \Phi^* \Phi + \frac{\lambda}{4!}(\Phi^* \Phi)^2,
\label{ham den7}
\end{equation}
where we neglected a time-dependent $c$-number term. Since it has
been shown in Sec. V that the coherent and coherent-thermal states
in the LvN approach give the identical result as the mean-field
and Hartree-Fock methods, we first divide the field into a
classical background and quantum fluctuations and then quantize
the latter according to the LvN approach. The field may be divided
into the homogeneous classical background field and quantum
fluctuations
\begin{equation}
\Phi ({\bf x}, t) = \phi_c (t) + \Phi_f ({\bf x}, t)
\end{equation}
such that the classical background is a coherent state of the
quantum field and the quantum fluctuations have symmetric states
such as the vacuum, number and thermal states:
\begin{equation}
\langle \hat{\Phi} \rangle = \phi_c (t),\quad \langle \hat{\Phi}_f
\rangle = 0. \label{sym st}
\end{equation}
The Hamiltonian density is then the sum of the classical
background, quantum fluctuations and interactions
\begin{equation}
{\cal H} ({\bf x}, t) = {\cal H}_c (t) + {\cal H}_f ({\bf x}, t) +
{\cal H}_{int} ({\bf x}, t) + \delta{\cal H}_{int} ({\bf x}, t),
\end{equation}
where
\begin{eqnarray}
{\cal H}_c (t) &=& \pi_{c}^2 + m^2 (t) \phi_c^2
+\frac{\lambda}{4!}\phi_c^4, \nonumber\\ {\cal H}_f ({\bf x}, t)
&=& \Pi_{f}^* \Pi_f + \nabla \Phi_f^* \cdot \nabla \Phi_f + m^2
(t)\Phi_f^* \Phi_f +\frac{\lambda}{4!}(\Phi_f^* \Phi_f)^2,
\nonumber\\ {\cal H}_{int} ({\bf x}, t) &=& \frac{\lambda}{3!}
\phi_c^2 \bigl(\Phi_f^* \Phi_f \bigr), \nonumber\\ \delta{\cal
H}_{int} ({\bf x}, t) &=& \pi_{c} \bigl(\Pi_{f}^* + \Pi_f \bigr) +
\bigl(\nabla \phi_c \bigr) \cdot \bigl(\nabla \Phi_f^* + \nabla
\Phi_f  \bigr) + m^2 (t) \phi_c \bigl(\Phi_f^* + \Phi_f \bigr)
\nonumber\\ && + \frac{\lambda}{4!} \Biggl\{\phi_c^2
\bigl(\Phi_f^{*2} + \Phi_f^2 \bigr) + 2\phi_c^3 \bigl(\Phi_f^* +
\Phi_f \bigr) \Bigr) + 2 \phi_c  \bigl(\Phi_f^* \Phi_f \bigr)
\bigl(\Phi_f^* + \Phi_f \bigr) \Biggr\}, \label{cl-fl7}
\end{eqnarray}
with $\pi_{c} = \dot{\phi}_c, \pi_{f} = \dot{\phi}_f$. As we are
interested in the symmetric state of quantum fluctuations
$\Phi_f$, so the expectation value of $\delta {\cal H}_{int}$
vanishes and will not be considered any more, since only the terms
involving $(\Phi_f^* \Phi_f)$ do not vanish when one takes the
expectation value with respect to the symmetric state.

Now the field $\Phi_f$ and $\Phi_f^*$ and their conjugate momenta
are decomposed into Fourier-modes according to Eqs.
(\ref{fourier1}) and (\ref{fourier2}). The Fourier transform of
the quartic term leads to
\begin{equation}
\int d^3 {\bf x} \bigl(\Phi_f^* \Phi_f \bigr)^2 = \prod_{j =
1}^{4} \int \frac{d^3 {\bf k}_j }{(2\pi)^3} \bigl(\phi_{{\bf
k}_1}^* \phi_{{\bf k}_2} \bigr) \bigl(\phi_{{\bf k}_3}^*
\phi_{{\bf k}_4} \bigr) \delta \bigl({\bf k}_1 + {\bf k}_3 - {\bf
k}_2 - {\bf k}_4 \bigr). \label{quartic}
\end{equation}
The symmetric state further restricts the integral in Eq.
(\ref{quartic}) to either $({\bf k}_1 = {\bf k}_2, {\bf k}_3 =
{\bf k}_4)$ or $({\bf k}_1 = {\bf k}_4, {\bf k}_2 = {\bf k}_3)$,
so Eq. (\ref{quartic}) leads to
\begin{equation}
\langle \int d^3 {\bf x} \bigl(\hat{\Phi}_f^* \hat{\Phi}_f
\bigr)^2 \rangle = 2 \Biggl[\int \frac{d^3 {\bf k} }{(2\pi)^3}
\langle \hat{\phi}_{\bf k}^* \hat{\phi}_{\bf k} \rangle \Biggr]^2
= 2 \langle \hat{\Phi}_f^* \hat{\Phi}_f \rangle^2.
\end{equation}
Keeping the symmetric state in mind, we find the the Fourier
transform of ${\cal H}_f + {\cal H}_{int}$
\begin{eqnarray}
{\cal H}_f + {\cal H}_{int} &=& \int \frac{d^3 {\bf k}}{(2\pi)^3}
\Biggl[\pi^*_{\bf k} (t) \pi_{\bf k} (t) + \Bigl( {\bf k}^2 + m^2
(t) + \frac{\lambda}{3!} \phi_c^2 (t) \Bigr) \phi^*_{\bf k} (t)
\phi_{\bf k} (t) \Biggr] \nonumber\\&& + \frac{\lambda}{2 \cdot
3!} \Biggl[\int \frac{d^3 {\bf k} }{(2\pi)^3} \phi_{\bf k}^* (t)
\phi_{\bf k} (t) \Biggr]^2. \label{ham mod}
\end{eqnarray}
Therefore the Hamiltonian for fluctuations is mode-decomposed into
the sum of infinite number of coupled anharmonic oscillators.

We apply the method for quantum anharmonic oscillators in Sec. V
to the Hamiltonian (\ref{ham mod}). According to the LvN approach,
we first introduce two pairs of the annihilation and creation
operators
\begin{eqnarray}
\hat{A}_{\bf k}  (t) = i \Bigl( \varphi^*_{\bf k} (t)
\hat{\pi}^*_{\bf k} - \dot{\varphi}^*_{\bf k} (t) \hat{\phi}_{\bf
k} \Bigr), \nonumber\\ \hat{A}_{\bf k}^{\dagger}  (t) = - i \Bigl(
\varphi_{\bf k} (t) \hat{\pi}_{\bf k} - \dot{\varphi}_{\bf k} (t)
\hat{\phi}^*_{\bf k} \Bigr), \label{pair1}
\end{eqnarray}
and
\begin{eqnarray}
\hat{A}_{\bf k}^*  (t) = i \Bigl( \varphi^*_{\bf k} (t)
\hat{\pi}_{\bf k} - \dot{\varphi}^*_{\bf k} (t) \hat{\phi}_{\bf
k}^* \Bigr), \nonumber\\ \hat{A}_{\bf k}^{*\dagger} (t) = - i
\Bigl( \varphi_{\bf k} (t) \hat{\pi}^*_{\bf k} -
\dot{\varphi}_{\bf k} (t) \hat{\phi}_{\bf k} \Bigr). \label{pair2}
\end{eqnarray}
We then require the creation and annihilation operators
(\ref{pair1}) to satisfy the LvN equation to obtain the equation
of motion
\begin{equation}
\ddot{\varphi}_{\bf k} (t) + \Biggl[ {\bf k}^2 + m^2 (t)  +
\frac{\lambda}{3!} \phi_c^2 (t) + \frac{\lambda}{3!} \int
\frac{d^3 {{\bf k}_1} }{(2\pi)^3} \hbar^2 \varphi^*_{{\bf k}_1}
(t) \varphi_{{\bf k}_1} (t) \Biggr] \varphi_{\bf k} (t) = 0,
\label{4 cl eq}
\end{equation}
where the expectation value is taken with respect to the vacuum
state, a symmetric state, as mentioned. The LvN equations for the
operators (\ref{pair2}) are the complex conjugate of Eq. (\ref{4
cl eq}).

The Fock space for each mode can be found similarly according to
Sec. V. We consider two symmetric states: the vacuum and thermal
states. The vacuum state that is annihilated by all $\hat{A}_{\bf
k}$ and $\hat{A}^*_{\bf k}$
\begin{equation}
\hat{A}_{\bf k} (t) \vert 0, t \rangle = 0, \quad \hat{A}^*_{\bf
k} (t) \vert 0, t \rangle = 0,
\end{equation}
leads to the expectation values
\begin{eqnarray}
\langle \hat{\Phi}^*_f \hat{\Phi}_f \rangle_{\rm V} = \int
\frac{d^3 {\bf k}}{(2\pi)^3} \Bigl[ \hbar^2 \varphi^*_{\bf k} (t)
\varphi_{\bf k} (t) \Bigr], \nonumber\\ \langle \hat{\Pi}^*_f
\hat{\Pi}_f \rangle_{\rm V} = \int \frac{d^3 {\bf k}}{(2\pi)^3}
\Bigl[ \hbar^2 \dot{\varphi}^*_{\bf k} (t) \dot{\varphi}_{\bf k}
(t) \Bigr],
\end{eqnarray}
The initial thermal state defined by the density operator for each
mode
\begin{eqnarray}
\hat{\rho}_{\rm T} (t) = \prod_{\bf k} \hat{\rho}_{\bf k} (t) =
\prod_{\bf k} \Biggl\{\frac{1}{Z_{\bf k}} \exp \Biggl[ - \beta
\hbar \omega_{i, {\bf k}} \Bigl(\hat{A}^{\dagger}_{\bf k} (t)
\hat{A}_{\bf k} (t) + \frac{1}{2} \Bigr) \Biggr] \nonumber\\
\times \frac{1}{Z^*_{\bf k}} \exp \Biggl[ - \beta \hbar \omega_{i,
{\bf k}} \Bigl(\hat{A}^{*\dagger}_{\bf k} (t) \hat{A}^*_{\bf k}
(t) + \frac{1}{2} \Bigr) \Biggr] \Biggr\}
\end{eqnarray}
leads to the expectation values
\begin{eqnarray}
\langle \hat{\Phi}^*_f \hat{\Phi}_f \rangle_{\rm T} &=&  {\rm Tr}
\Bigl[\hat{\rho}_{\rm T} (t) \hat{\Phi}^*_f \hat{\Phi}_f \Bigr] =
\int \frac{d^3 {\bf k}}{(2\pi)^3} \Bigl[ \hbar^2 \varphi^*_{\bf k}
(t) \varphi_{\bf k} (t) \coth (\frac{\beta \hbar \omega_{i, {\bf
k}}}{2}) \Bigr], \nonumber\\ \langle \hat{\Pi}^*_f \hat{\Pi}_f
\rangle_{\rm T} &=& {\rm Tr} \Bigl[\hat{\rho}_{\rm T} (t)
\hat{\Phi}^*_f \hat{\Phi}_f \Bigr] = \int \frac{d^3 {\bf
k}}{(2\pi)^3} \Bigl[ \hbar^2 \dot{\varphi}^*_{\bf k} (t)
\dot{\varphi}_{\bf k} (t) \coth (\frac{\beta \hbar \omega_{i, {\bf
k}}}{2}) \Bigr]. \label{4 ex}
\end{eqnarray}
Then the effective Hamiltonian for the classical background with
all the contributions from fluctuations in the initial thermal
state is given by
\begin{eqnarray}
H_C (t) &\equiv& \hat{\cal H}_c  + \langle \hat{\cal H}_{int}
\rangle_{\rm T} \nonumber\\ &=& \pi_{c}^2 + \Biggl[ m^2 (t) +
\frac{\lambda}{3!} \int \frac{d^3 {\bf k}}{(2\pi)^3} \hbar^2
\varphi^*_{\bf k} (t) \varphi_{\bf k} (t) \coth (\frac{\beta \hbar
\omega_{i, {\bf k}}}{2}) \Biggr] \phi_c^2
+\frac{\lambda}{4!}\phi_c^4, \label{4 ham1}
\end{eqnarray}
and that for fluctuations by
\begin{eqnarray}
H_{\rm F} &\equiv& \langle \hat{\cal H}_f + \hat{\cal H}_{int}
\rangle_{\rm T} \nonumber\\ &=& \int \frac{d^3 {\bf k}}{(2\pi)^3}
\Biggl[\hbar^2 \dot{\varphi}^*_{\bf k} (t) \dot{\varphi}_{\bf k}
(t) + \Bigl( {\bf k}^2 + m^2 (t)  + \frac{\lambda}{3!} \phi_c^2
(t) \Bigr) \hbar^2 \varphi^*_{\bf k} (t) \varphi_{\bf k} (t)
\Biggr] \coth (\frac{\beta \hbar \omega_{i, {\bf k}}}{2})
\nonumber\\&& + \frac{\lambda}{2 \cdot 3!} \Biggl[\int \frac{d^3
{\bf k}}{(2\pi)^3} \hbar^2 \varphi^*_{\bf k} (t) \varphi_{\bf k}
(t) \coth (\frac{\beta \hbar \omega_{i, {\bf k}}}{2}) \Biggr]^2.
\label{4 ham2}
\end{eqnarray}
The effective Hamiltonian from the initial vacuum state is the
zero-temperature limit $(\beta \rightarrow \infty)$, {\it i.e.},
$\coth (\beta \hbar \omega_{i, {\bf k}}/2) \rightarrow 1$.

On the other hand, in the effective action method of Sec. V.B, the
$\varphi_{\bf k}$ is a parameter that will be determined by the
Hamilton equations. By writing $\varphi_{\bf k}$ in the polar form
\begin{equation}
\varphi_{\bf k} = \frac{\zeta_{\bf k} (t)}{\sqrt{\hbar}} e^{- i
\theta_{\bf k}},
\end{equation}
and by introducing $p_{\zeta_{\bf k}} = \dot{\zeta}_{\bf k}$, the
effective Hamiltonian for the ${\bf k}$-mode (\ref{4 ham2}) can be
rewritten as
\begin{equation}
H_{{\rm T},{\bf k}} (t) =  \hbar \coth (\frac{\beta \hbar
\omega_{i, {\bf k}}}{2}) \Biggl[p^2_{\zeta_{\bf k}} + \omega_{\bf
k}^2 (t) \zeta^2_{\bf k} + \frac{1}{8 \zeta^2_{\bf k}}\Biggr] =
\coth (\frac{\beta \hbar \omega_{i, {\bf k}}}{2}) H_{{\rm V}, {\bf
k}} (t). \label{4 ham3}
\end{equation}
where
\begin{equation}
\omega_{\bf k}^2 (t) =  {\bf k}^2 + m^2 (t) + \frac{\lambda}{3!}
\phi_c^2 (t) + \frac{\lambda \hbar}{3!} \int \frac{d^3 {\bf
k}}{(2\pi)^3} \zeta^2_{\bf k} (t) \coth (\frac{\beta \hbar
\omega_{i, {\bf k}}}{2}). \label{4 freq}
\end{equation}
The Hamilton equations are
\begin{eqnarray}
\frac{d \zeta_{\bf k}}{dt} &=& \frac{\partial}{\partial
p_{\zeta_{\bf k}}} \Bigl[ \frac{H_{{\rm T},{\bf k}} (t)}{\hbar
\coth (\frac{\beta \hbar \omega_{i, {\bf k}}}{2})}\Bigr] =
p_{\zeta_{\bf k}}, \nonumber\\ \frac{d p_{\zeta_{\bf k}}}{dt} &=&
- \frac{\partial}{\partial \zeta_{\bf k}} \Bigl[ \frac{H_{{\rm
T},{\bf k}} (t)}{\hbar \coth (\frac{\beta \hbar \omega_{i, {\bf
k}}}{2})}\Bigr] = - \omega^2_{\bf k} (t) \zeta_{\bf k} +
\frac{1}{4 \zeta_{\bf k}^3}. \label{4 ham eq}
\end{eqnarray}
These Hamilton equations are identical to Eq. (\ref{4 cl eq}),
because the effective action method is equivalent to the LvN
approach as shown in Sec. V.B. The Hamilton equations of motion
for the classical background are given by
\begin{equation}
\ddot{\phi}_c (t) + \Biggl[ m^2 (t) + \frac{\lambda}{3!} \int
\frac{d^3 {\bf k}}{(2\pi)^3} \hbar^2 \varphi^*_{\bf k} (t)
\varphi_{\bf k} (t) \coth (\frac{\beta \hbar \omega_{i, {\bf
k}}}{2}) \Biggr] \phi_c (t) +\frac{\lambda}{2 \cdot 3!}\phi_c^3 =
0. \label{4 cl eq2}
\end{equation}

The different quantum states given in Sec. V can be chosen to
describe the different processes for the phase transition. We
assume that the system starts from either the thermal or
coherent-thermal states before the quench and evolves according to
the functional Schr\"{o}dinger equation during and after the
quench. In this sense the process is completely determined once
the initial quantum state is prescribed. In the first case of the
symmetric thermal state, each mode has the zero expectation value
for the position and momentum, but its dynamics is still governed
by Eq. (\ref{4 cl eq}) with $\phi_c = 0$. In the second case of
the coherent-thermal state, the classical field $\phi_c$, which is
a coherent state of the homogeneous field, plays the role of an
order parameter and is influenced by thermal fluctuations of
$\phi_{\bf k}$. As the first case is a limit of the second one, we
shall first focus on the coherent-thermal and then obtain the
thermal state result by taking the limit $\phi_c (t) =
\dot{\phi}_c (t) = 0$ in the end.

It is very difficult to solve analytically the equations of motion
(\ref{4 cl eq}) and (\ref{4 cl eq2}). At best we have to rely on
the adiabatic solutions that can be found in some important
physical regimes. We assume that far before the quench $(t
\rightarrow - \infty)$ the system starts from a thermal
fluctuations around the order parameter with
\begin{equation}
\phi_c (- \infty) \approx 0, \quad \dot{\phi}_c (- \infty) \approx
0. \label{therm fluc}
\end{equation}
As $\phi_c$ is a classical field, the uncertainty principle does
not prohibit us from even taking the limit $\phi_c (- \infty) =
\dot{\phi}_c (- \infty) = 0$, which leads to the symmetric thermal
state. The initial data for $\phi_c$ take very small values with
high probability by some distribution function. For the sake of
simplicity we consider the instantaneous quench first. Before the
quench time, $m^2 (t)$ takes the initial constant value $m^2_i$
and $\phi_c$ remains close to zero, so $\omega_{\bf k}$ changes
very little during the evolution. The solutions to Eq. (\ref{4 cl
eq}) are approximately given by
\begin{equation}
\varphi_{i, {\bf k}} (t) = \frac{1}{\sqrt{2 \hbar \Omega_{i,{\bf
k}}}} e^{- i \Omega_{i, {\bf k}} t}, \label{4 in sol}
\end{equation}
and Eq. (\ref{4 freq}) yields the gap equation
\begin{equation}
\Omega^2_{i, {\bf k}} = m^2_i + k^2 + \frac{\lambda \hbar}{3!}
\int \frac{d^3 {\bf k}}{(2\pi)^3} \frac{1}{2 \Omega_{i, {\bf k}}}
\coth (\frac{\beta \hbar \omega_{i, {\bf k}}}{2}). \label{4 freq2}
\end{equation}
The infinite quantity that appears in Eq. (\ref{4 freq2}) will be
absorbed by the bare coupling parameters $m^2_i$ and $\lambda$ to
result in the renormalized ones and a finite equation for
$\Omega^2_{i, {\bf k}} \cite{chang,boyanovsky}$. The solutions
(\ref{4 in sol}) hold till the quench time $t = 0$.

But after the quench time, $m^2 (t)$ changes to $- m_f^2$, and
therefore, the long wavelength modes become unstable and the
spinodal instability begins. The long wavelength solutions are
given by
\begin{equation}
\varphi_{f_{U}, {\bf k}} (t) = \frac{1}{\sqrt{2 \hbar \Omega_{i,
{\bf k}}}} \Bigl[- i \frac{\Omega_{i, {\bf k}}}{\tilde{\Omega}_{f,
{\bf k}} (t)} \sinh (\int_{0}^{t} \tilde{\Omega}_{f, {\bf k}}(t))
+ \cosh (\int_{0}^{t} \tilde{\Omega}_{f, {\bf k}} (t)) \Bigr],
\end{equation}
and the short wavelength solutions by
\begin{equation}
\varphi_{f_{S}, {\bf k}} (t) = \frac{1}{\sqrt{2 \hbar \Omega_{i,
{\bf k}}}} \Bigl[- i \frac{\Omega_{i, {\bf k}}}{\Omega_{f, {\bf
k}}(t)} \sin (\int_{0}^{t} \Omega_{f, {\bf k}}(t)) + \cos
(\int_{0}^{t} \Omega_{f, {\bf k}}(t)) \Bigr],
\end{equation}
where
\begin{eqnarray}
\tilde{\Omega}^2_{f, {\bf k}} (t) &=& m_f^2 - k^2 -
\frac{\lambda}{3!} \phi_c^2 \nonumber\\&& - \frac{\lambda
\hbar}{3!} \int_{0}^{k_{\Lambda}} \frac{d^3 {\bf k}}{(2\pi)^3}
\frac{\coth (\frac{\beta \hbar \omega_{i, {\bf k}}}{2})}{2
\Omega_{i, { \bf k}}} \Biggl[\Biggl\{ \Bigl(\frac{\Omega_{i, {\bf
k}}}{\tilde{\Omega}_{f, {\bf k}}(t)} \Bigr)^2 + 1 \Biggr\}^2 \sinh
(\int_{0}^{t} \tilde{\Omega}_{f, {\bf k}}(t)) + 1 \Biggr]
\nonumber\\ && - \frac{\lambda \hbar}{3!}
\int_{k_{\Lambda}}^{\infty} \frac{d^3 {\bf k}}{(2\pi)^3}
\frac{\coth (\frac{\beta \hbar \omega_{i, {\bf k}}}{2})}{2
\Omega_{i, { \bf k}}} \Biggl[\Biggl\{ \Bigl(\frac{\Omega_{i, {\bf
k}}}{\Omega_{f, {\bf k}}(t)} \Bigr)^2 - 1 \Biggr\}^2 \sin
(\int_{0}^{t} \Omega_{f, {\bf k}}(t)) + 1 \Biggr], \label{4 freq5}
\end{eqnarray}
and $\Omega^2_{f, {\bf k}} (t) = - \tilde{\Omega}^2_{f, {\bf k}}
(t)$. In contrast with the free scalar model in Sec. VI, not only
the duration of the instability but also the band of unstable
modes decrease in time due to the back-reaction $\hbar^2
\varphi^*_{{\bf k}}(t) \varphi_{{\bf k}}(t)$ of the
self-interaction and $\phi_c^2 (t)$ of the classical background.
The classical background obeying Eq. (\ref{4 cl eq2}), though it
remained around the initial vacuum under thermal fluctuations
before the quench time, begins to roll down from the false vacuum
to the true vacuum at first largely due to the exponentially
growing unstable modes and then to the self-interaction. The
competition between $- m_f^2$ and $\langle \hat{\Phi}_f^*
\hat{\Phi}_f \rangle$ together $\phi_c^2$ determines the spinodal
line, beyond which the instability stops and the fluctuations
begin to oscillate around the true vacuum. Therefore the
self-interacting phase transition model has a natural exit to the
spinodal instability \cite{cormier2}.

We now turn to the case of symmetric thermal state. Once the
initial condition is prescribed such that $\phi_c (- \infty) =
\dot{\phi}_c (- \infty) = 0$, the classical field $\phi_c (t)$
remains in the false vacuum even during the quench and has the
trivial solution $\phi_c (t) = 0$ for all times. Hence Eq. (\ref{4
cl eq2}) is identically satisfied, and the dynamics of the second
order phase transition is entirely governed by Eq. (\ref{4 cl eq})
for quantum fluctuations, which extends the free scalar field
model considered on Sec. VI. Though each mode of quantum
fluctuations has the zero expectation value, the Wigner function
becomes sharply peaked around its classical trajectory as the
quench proceeds \cite{kim-lee}. Even without the classical field
$\phi_c$ the quantum contribution from the self-interaction in Eq.
(\ref{4 cl eq}) still prevents the unstable modes from growing
indefinitely and provides a natural exit to the spinodal
instability. This implies that the quantum dynamics of the second
order phase transition is classically correlated and exhibits most
of the essential points described by the order parameter under
thermal fluctuations.

Finally we comment on the formation process of topological
defects. The topological defects formed from the second order
phase transition in this paper are domain walls. The correlation
of domain walls can be determined by the two-point thermal
correlation function
\begin{equation}
G_{\rm T} ({\bf y}, {\bf x}, t) = \langle \hat{\Phi}^* ({\bf y},
t) \hat{\Phi} ({\bf x}, t) \rangle_{\rm T} = \int
\frac{d^3k}{(2\pi)^3} \Bigl[ \hbar^2 \varphi^*_{\bf k} (t)
\varphi_{\bf k} (t) \coth (\frac{\beta \hbar \omega_{i, {\bf
k}}}{2}) \Bigr] e^{i {\bf k} \cdot ({\bf x} - {\bf y})},
\end{equation}
where $\omega_{i, {\bf k}} = \omega_{\bf k} (t = - \infty)$. After
the quench, the two-point correlation function is dominated by the
unstable modes $\varphi_{f_U}$. The scaling behavior of two-point
correlation function in Sec. VI holds still before reaching the
spinodal line. That is, the size of domains grows according to the
power law $t^{1/4}$ during the quench and the Cahn-Allen relation
$t^{1/2}$ after the completion of quench. However, the domains can
not grow indefinitely due to the back-reaction of the
self-interaction and the classical background. As each unstable
mode reaches further the spinodal line, its solution stops
exponential growing and oscillates. The behavior of the two-point
thermal correlation function changes from that in Sec. VI, and one
would expect a different scaling relation for the domain size,
probably with a small power than the Cahn-Allen relation.

\section{Conclusion}

In this paper we have elaborated the recently introduced
Liouville-von Neumann (LvN) approach to describe properly the
time-dependent nonequilibrium systems. The systems interacting
directly with environments or undergoing phase transitions are
such nonequilibrium systems. These systems are characterized by
time-dependent coupling parameters and their true nonequilibrium
evolution deviates significantly from the equilibrium one when
their coupling parameters differ greatly from their initial
values. In this case the systems evolve completely out of
equilibrium. For that purpose there have been developed many
different methods such as the closed time-path integral method,
sometimes in conjunction with the large $N$-expansion, mean-field,
Hartree-Fock method. The LvN approach developed in this paper is a
canonical method that unifies the functional Schr\"{o}dinger
equation for the quantum evolution of pure states and the LvN
equation for the quantum description of mixed states of either
equilibrium or nonequilibrium. Because the LvN approach shares all
the useful techniques with quantum mechanics and quantum
many-particle systems, it turns out to be a powerful method for
describing time-dependent harmonic oscillators and anharmonic
oscillators, and provides a rigorous and systematic method for
describing time-dependent phase transitions.

By applying the LvN approach to time-dependent harmonic
oscillators, we have found exactly the nonequilibrium quantum
evolution evolving from various initial states such as the vacuum,
number, coherent and thermal states. In this case the LvN approach
is based on two operators, the so-called the annihilation and
creation operators, that satisfy the quantum LvN equation, so it
is straightforward to construct the Fock space of number states
and the density operator according to the standard technique of
quantum mechanics. We have thus obtained the density operator in
terms of the classical solution, and by using the exact wave
functions for number states, have been able to find the explicit
form of the density matrix. In particular, the density matrix
provides us with a criterion on nonequilibrium vs. equilibrium
evolution. Moreover, the LvN approach has been applied to the
time-dependent inverted harmonic oscillators, which can be
regarded as models for second order phase transitions. For
time-dependent anharmonic oscillators, we have found approximately
the nonequilibrium evolution of the symmetric Gaussian, coherent
and thermal states at the lowest order of the coupling constant of
the quartic term. It has been shown that the LvN approach is
equivalent to the effective action method and to the mean field or
Hartree-Fock method.

Finally we have applied the LvN approach to the systems undergoing
the symmetry breaking second order phase transition. In
particular, due to the quench the coupling parameters change the
sign during the evolution. As field models we have studied a free
massive scalar field with an instantaneous and a finite smooth
quench. By applying the LvN approach to this symmetry breaking
system we have found the two-point vacuum and thermal correlation
functions. It has proved that the spinodal instability leads to
the $t^{1/4}$-scaling relation for domain sizes during the quench
and the classical Cahn-Allen relation after the completion of
quench. The Cahn-Allen scaling relation confirms the result for
the instantaneous quench model in Refs. \cite{boyanovsky,bowick}.
One prominent feature of the finite smooth quench model is the
time-lag occurring at the cubic power of the quench period in the
Cahn-Allen scaling relation after the completion of quench. The
inclusion of a self-interacting term shuts off the spinodal
instability after crossing the spinodal line and gives rise to a
natural exit for the spinodal decomposition. Not treated in detail
in this paper is the very slow quench effect, which may show a
transient resonance of decaying solution of long wavelength modes
and will be addressed in a future research.

\acknowledgments

We would like to thank S. G. Kim for many useful discussions. This
work was supported by the KOSEF under Grant No. 1999-2-112-003-5.

\appendix
\section{Harmonic Oscillator Wave Functions}

The vacuum state of the time-dependent oscillator is annihilated
by $\hat{a}(t)$:
\begin{equation}
\hat{a} (t) \vert 0, t \rangle = 0. \label{a1}
\end{equation}
In the coordinate representation
\begin{equation}
\Psi_0 (q, t) = \langle q \vert 0, t \rangle,
\end{equation}
Eq. (\ref{a1}) becomes
\begin{equation}
i \Biggl[u^* \frac{\hbar}{i} \frac{\partial}{\partial q} - m
\dot{u}^* q \Biggr] \Psi_0 (q, t) = 0.
\end{equation}
We thus obtain the normalized wave function for the vacuum state
\begin{equation}
\Psi_0 (q, t) = \Biggl(\frac{1}{2 \pi \hbar^2 u^* u} \Biggr)^{1/4}
\exp \Biggl[\frac{i}{2}\frac{m}{\hbar} \frac{\dot{u}^*}{u^*} q^2
\Biggr]. \label{a2}
\end{equation}
The wave function for the $n$th number state is obtained by
applying the creation operator $\hat{a}^{\dagger} (t)$ $n$-times
\begin{equation}
\Psi_n (q, t) = \frac{1}{\sqrt{n!}} \Bigl(\hat{a}^{\dagger} (t)
\Bigr)^{n} \Psi_0 (q, t).
\end{equation}
By making use of the relation
\begin{equation}
\Bigl(\hat{a}^{\dagger} (t) \Bigr)^n \Psi (q,t) = \Bigl(- \hbar u
\Bigr)^n e^{\frac{i}{2}\frac{m}{\hbar} \frac{\dot{u}}{u} q^2}
\Bigl(\frac{\partial}{\partial q} \Bigr)^n \Biggl(e^{-
\frac{i}{2}\frac{m}{\hbar} \frac{\dot{u}}{u} q^2} \Psi(q,
t)\Biggr),
\end{equation}
and the definition of the Hermite polynomial
\begin{equation}
H_n (x) = (-1)^n e^{x^2} \Bigl(\frac{d}{dx}\Bigr)^n e^{- x^2},
\end{equation}
we obtain the wave function
\begin{equation}
\Psi_n (q, t) = \Biggl(\frac{1}{2 \pi \hbar^2 u^* u} \Biggr)^{1/4}
\frac{1}{\sqrt{2^n n!}} \Biggl(\frac{u}{u^*} \Biggr)^n H_n (x)
\exp \Biggl[\frac{i}{2}\frac{m}{\hbar} \frac{\dot{u}^*}{u^*} q^2
\Biggr], \label{a3}
\end{equation}
where
\begin{equation}
x = \frac{q}{\sqrt{2 \hbar^2 u^*u}}.
\end{equation}
Here we have also used the wronskian (\ref{boun con}).

\section{Density Matrix}

In the coordinate representation, the density operator defined by
\begin{equation}
\hat{\rho}_{\rm T} (t) = \frac{1}{Z_N}e^{- \beta \hbar \omega_0
(\hat{N} (t) + \frac{1}{2})}, \label{b1}
\end{equation}
where $Z_N$ is the partition function, becomes
\begin{equation}
\rho_{\rm T} (q', q, t) = \frac{1}{Z_N} \sum_{n = 0}^{\infty}
\Psi_n (q', t) \Psi^*_n (q, t) e^{- \beta \hbar \omega_0 (n +
\frac{1}{2})}. \label{b2}
\end{equation}
By substituting (\ref{a3}) into (\ref{b2}), we obtain
\begin{equation}
\rho_{\rm T} (q', q, t) = \frac{1}{Z_N} \Biggl(\frac{1}{2 \pi
\hbar^2 u^*u} \Biggr)^{1/2} \sum_{n = 0}^{\infty} \frac{1}{2^n n!}
H_n (x') H_n (x) e^{- \beta \hbar \omega_0 (n + \frac{1}{2})}
e^{\frac{i}{2}\frac{m}{\hbar} \frac{\dot{u}^*}{u^*} q'^2 -
\frac{i}{2}\frac{m}{\hbar} \frac{\dot{u}}{u} q^2}.
\end{equation}
Following Kubo's method \cite{kubo}, we rewrite the product of
Hermite polynomials as \cite{gradshteyn}
\begin{equation}
H_n (x') H_n (x) = \frac{1}{\pi} e^{x'^2 + x^2} \int_{-
\infty}^{\infty} \int_{- \infty}^{\infty} dz_1 dz_2 (2iz_1)^n (2 i
z_2)^n e^{- z_1^2 - 2 i x' z_1 - z_2^2 - 2 i x z_2}
\end{equation}
and sum over $n$ to obtain
\begin{eqnarray}
\rho_{\rm T} (q', q, t) &=& \frac{1}{Z_N} \Biggl(\frac{1}{2 \pi
\hbar^2 u^*u} \Biggr)^{1/2} \frac{e^{- \frac{\beta \hbar
\omega_0}{2}}}{\pi} \exp \Biggl[ (x'^2 + x^2) +
\frac{i}{2}\frac{m}{\hbar} \Biggl\{ \frac{\dot{u}^*}{u^*} q'^2 -
\frac{\dot{u}}{u} q^2 \Biggr\} \Biggr] \nonumber\\ && \times
\int_{- \infty}^{\infty} \int_{- \infty}^{\infty} dz_1 dz_2 \exp
\Biggl[- z_1^2 - 2 i x' z_1 - z_2^2 - 2 i x z_2 - 2 z_1 z_2 e^{-
\beta \hbar \omega_0} \Biggr].
\end{eqnarray}
After doing the integral and using the identity
\begin{equation}
x^2 = \frac{i}{2} \frac{m}{\hbar} \Bigl(\frac{\dot{u}}{u} -
\frac{\dot{u}^*}{u^*} \Bigr) q^2,
\end{equation}
we finally obtain
\begin{eqnarray}
\rho_{\rm T} (q', q, t) &=& \frac{1}{Z_N} \Biggl[\frac{1}{4 \pi
\hbar^2 u^* u \sinh(\beta \hbar \omega_0)}\Biggr]^{1/2}  \exp
\Biggl[ \frac{i}{2} \hbar m \frac{d}{dt} \ln (u^* u) (x'^2 - x^2)
\Biggr] \nonumber\\ && \times \exp \Biggl[-\frac{1}{4} \Biggl\{
(x' + x)^2 \tanh(\frac{\beta \hbar \omega_0}{2}) + (x'-x)^2
\coth(\frac{\beta \hbar \omega_0}{2}) \Biggr\} \Biggr]. \label{b3}
\end{eqnarray}
In the special case of the time-independent oscillator we recover
the classical result by Kubo \cite{kubo} by substituting the
complex solution
\begin{equation}
u(t) = \frac{e^{- i \omega_0 t}}{\sqrt{2 \hbar m \omega_0}}
\end{equation}
into Eq. (\ref{b3}).

\section{Steepest Decent Method}

The integral appearing in the two-point correlation function has
the form
\begin{equation}
I = \int dx x e^{-\gamma x^2} \sin(yx) F(x) \label{int}
\end{equation}
where $F(x)$ is a slowly varying function. Note that $xe^{- \gamma
x^2}$ is a highly peaked function that varies rapidly. We let
\begin{equation}
 e^{g(x)} \equiv xe^{- \gamma x^2}
\end{equation}
where
\begin{equation}
g(x) = - \gamma x^2 + \ln (x).
\end{equation}
We expand $g(x)$ in a Taylor series and truncate it up to the
quadratic term around the maximum point $x_0 = \frac{1}{\sqrt{2
\gamma}}$
\begin{equation}
g (x) \approx g(x_0) - 2\gamma (x - x_0)^2,
\end{equation}
and rewrite the integrand as
\begin{eqnarray}
x e^{- \gamma x^2} \sin(yx) F(x) &\approx& x_0e^{-\gamma x_0^2}
F(x_0) e^{- 2\gamma (x-x_0)^2} \frac{e^{iyx} - e^{-iyx}}{2i}
\nonumber\\&=&  x_0e^{-\gamma x_0^2} F(x_0) \Biggl\{ \exp\Biggl[-
2\gamma \Bigl(x-x_0 - i \frac{y}{4 \gamma}\Bigr)^2\Biggr]
\frac{e^{i y x_0}}{2i} \nonumber\\ && + \exp\Biggl[- 2\gamma
\Bigl(x-x_0 +i \frac{y}{4 \gamma}\Bigr)^2 \Biggr] \frac{(-1) e^{-i
y x_0}}{2i} \Biggr\}e^{ - \frac{y^2}{8 \gamma}}.
\end{eqnarray}
The Gaussian integrals contribute equally, so we obtain
\begin{equation}
I \approx \Biggl(\frac{\pi}{4 \gamma} \Biggr)^{1/2} x_0e^{-\gamma
x_0^2} F(x_0) \sin(yx_0) e^{ - \frac{y^2}{8 \gamma}}.
\end{equation}

\section{Asymptotic Form for Unstable Modes in a Finite Quench}

To find the contribution to the two-point functions from the
unstable growing modes for a finite quench $(\tilde{\omega}_{f,
{\bf k}} \tau < 1)$, we need to evaluate $\Gamma
(\tilde{\omega}_{f, {\bf k}})$ and
$\Gamma(\frac{\tau}{2}(\tilde{\omega}_{f, {\bf k}}- i \omega_{i,
{\bf k}}))$. These gamma functions are rewritten as
\begin{eqnarray}
\Gamma (\tilde{\omega}_{f, {\bf k}} \tau) &=& \frac{\Gamma (1+
\tilde{\omega}_{f, {\bf k}} \tau)}{\tilde{\omega}_{f, {\bf k}}
\tau}, \nonumber\\ \Gamma(\frac{\tau}{2}(\tilde{\omega}_{f, {\bf
k}}- i \omega_{i, {\bf k}})) &=& \frac{\Gamma(1+
\frac{\tau}{2}(\tilde{\omega}_{f, {\bf k}}- i \omega_{i, {\bf
k}}))}{\frac{\tau}{2}(\tilde{\omega}_{f, {\bf k}}- i \omega_{i,
{\bf k}})}.
\end{eqnarray}
We further make use of the expansion formula \cite{abramowitz}
\begin{equation}
\ln \Gamma (1 + z) = - \ln (1+z) + (1 - \gamma) z + \sum_{n =
2}^{\infty}(-1)^n \Bigl[\zeta(n) -1 \Bigr]\frac{z^n}{n}, \quad
(|z| < 2), \label{gamma ex}
\end{equation}
where
\begin{equation}
\gamma = \lim_{m \rightarrow \infty} \Biggl[\sum_{k = 1}^{m}
\frac{1}{k} - \ln (m) \Biggr] = 0.5772156649 \cdots
\end{equation}
is the Euler's constant and $\zeta(n)$ is the Riemann Zeta
function
\begin{equation}
\zeta (n) = \sum_{k = 1}^{\infty} \frac{1}{k^n}.
\end{equation}
Here
\begin{equation}
z = \tilde{\omega}_{f, {\bf k}} \tau, \quad z =
\frac{\tau}{2}(\tilde{\omega}_{f, {\bf k}} - i \omega_{i, {\bf k}}
\tau).
\end{equation}
We now expand the gamma functions up to the cubic power of $\tau$:
\begin{eqnarray}
\Gamma (1+ \tilde{\omega}_{f, {\bf k}} \tau) &=& \frac{1}{1 +
\tilde{\omega}_{f, {\bf k}} \tau} \exp \Biggl[ (1 - \gamma)
(\tilde{\omega}_{f, {\bf k}} \tau) + \frac{1}{2} [\zeta (2) -1]
(\tilde{\omega}_{f, {\bf k}} \tau)^2 \nonumber\\ &-& \frac{1}{3}
[\zeta(3) -1] (\tilde{\omega}_{f, {\bf k}} \tau)^3 + {\cal O}
(\tau^4) \Biggr]
\end{eqnarray}
and
\begin{eqnarray}
\Bigl| \Gamma(1+ \frac{\tau}{2}(\tilde{\omega}_{f, {\bf k}}- i
\omega_{i, {\bf k}})) \Bigr|^2 &=& \frac{1}{|1 +
\frac{\tau}{2}(\tilde{\omega}_{f, {\bf k}}- i \omega_{i, {\bf k}})
|^2} \exp \Biggl[(1 - \gamma) (\tilde{\omega}_{f, {\bf k}} \tau)
\nonumber\\ &+& \frac{1}{2} [ \zeta (2) -1] (\frac{\tau}{2})^2
\Bigl\{ (\tilde{\omega}_{f, {\bf k}}- i \omega_{i, {\bf k}})^2 +
(\tilde{\omega}_{f, {\bf k}}+ i \omega_{i, {\bf k}})^2 \Bigr\}
\nonumber\\ &-&  \frac{1}{3} [ \zeta (3) -1] (\frac{\tau}{2})^3
\Bigl\{ (\tilde{\omega}_{f, {\bf k}}- i \omega_{i, {\bf k}})^3 +
(\tilde{\omega}_{f, {\bf k}}- i \omega_{i, {\bf k}})^3 \Bigr\} +
{\cal O} (\tau^4) \Biggr].
\end{eqnarray}
Therefore it follows that
\begin{eqnarray}
&&\Biggl|\frac{\Gamma (1+ \tilde{\omega}_{f, {\bf k}}
\tau)}{\Gamma^2(1+ \frac{\tau}{2}(\tilde{\omega}_{f, {\bf k}}- i
{\omega}_{i, {\bf k}}))} \Biggr|^2 = \Biggl[ \frac{\Bigl(1 +
\frac{\tau}{2} \tilde{\omega}_{f, {\bf k}} \Bigr)^2 +
\frac{\tau^2}{4} \omega_{i, {\bf k}}^2 }{1 + \tilde{\omega}_{f,
{\bf k}}\tau} \Biggr]^2
 \nonumber\\ && \quad \times \exp \Biggl[\frac{1}{2}[ \zeta(2) -1] \tau^2
\Bigl(\tilde{\omega}_{f, {\bf k}}^2 + \omega^2_{i, {\bf k}} \Bigr)
- \frac{1}{4}[ \zeta(3) -1] \tau^3 \tilde{\omega}_{f, {\bf k}}
\Bigl(\tilde{\omega}_{f, {\bf k}}^2 + \omega^2_{i, {\bf k}} \Bigr)
+ {\cal O} (\tau^4)  \Biggr].
\end{eqnarray}
We finally get
\begin{eqnarray}
&&\Biggl|\frac{\Gamma (\tilde{\omega}_{f, {\bf k}}
\tau)}{\Gamma^2( \frac{\tau}{2}(\tilde{\omega}_{f, {\bf k}}- i
{\omega}_{i, {\bf k}}))} \Biggr|^2 = \Biggl[\frac{\frac{\tau^2}{4}
\Bigl(\tilde{\omega}_{f, {\bf k}}^2 + \omega_{i, {\bf k}}^2 \Bigr)
}{\tilde{\omega}_{f, {\bf k}} \tau } \Biggr]^2 \Biggl[
\frac{\Bigl(1 + \frac{\tau}{2} \tilde{\omega}_{f, {\bf k}}
\Bigr)^2 + \frac{\tau^2}{4} \omega_{i, {\bf k}}^2 }{1+
\tilde{\omega}_{f, {\bf k}}\tau} \Biggr]^2
 \nonumber\\ && \quad \times \exp \Biggl[\frac{1}{2}[ \zeta(2) -1] \tau^2
\Bigl(\tilde{\omega}_{f, {\bf k}}^2 + \omega^2_{i, {\bf k}} \Bigr)
- \frac{1}{4}[ \zeta(3) -1] \tau^3 \tilde{\omega}_{f, {\bf k}}
\Bigl(\tilde{\omega}_{f, {\bf k}}^2 + \omega^2_{i, {\bf k}} \Bigr)
+ {\cal O} (\tau^4)  \Biggr].
\end{eqnarray}
The last relation necessary for the two-point correlation function
is \cite{abramowitz}
\begin{equation}
\Bigl|\Gamma(1 - i \omega_{i, {\bf k}}\tau ) \Bigr|^2 = \frac{\pi
\omega_{i, {\bf k}}\tau}{\sinh(\pi \omega_{i, {\bf k}}\tau)}.
\end{equation}

\end{document}